\documentclass[11pt]{pmet}

\submit

\RequirePackage{natbib}
\usepackage{amsmath,amssymb}

\usepackage{multirow}
\usepackage{threeparttable}

\makeatletter
\@ifundefined{MPFourScale}{\def\MPFourScale{1.00}}{}
\DeclareFontFamily{U}{mp4}{}%
\DeclareFontShape{U}{mp4}{m}{n}{<->s * [\MPFourScale]cmb10}{}
\DeclareSymbolFont{boldgreekuc}{U}{mp4}{m}{n}
\DeclareMathSymbol{\bfAlpha}{\mathord}{boldgreekuc}{"41}
\DeclareMathSymbol{\bfBeta}{\mathord}{boldgreekuc}{"42}
\DeclareMathSymbol{\bfPsi}{\mathord}{boldgreekuc}{"09}
\DeclareMathSymbol{\bfDelta}{\mathord}{boldgreekuc}{"01}
\DeclareMathSymbol{\bfEpsilon}{\mathord}{boldgreekuc}{"45}
\DeclareMathSymbol{\bfPhi}{\mathord}{boldgreekuc}{"08}
\DeclareMathSymbol{\bfGamma}{\mathord}{boldgreekuc}{"00}
\DeclareMathSymbol{\bfEta}{\mathord}{boldgreekuc}{"48}
\DeclareMathSymbol{\bfIota}{\mathord}{boldgreekuc}{"49}
\DeclareMathSymbol{\bfXi}{\mathord}{boldgreekuc}{"04}
\DeclareMathSymbol{\bfKappa}{\mathord}{boldgreekuc}{"4B}
\DeclareMathSymbol{\bfLambda}{\mathord}{boldgreekuc}{"03}
\DeclareMathSymbol{\bfMu}{\mathord}{boldgreekuc}{"4D}
\DeclareMathSymbol{\bfNu}{\mathord}{boldgreekuc}{"4E}
\DeclareMathSymbol{\bfPi}{\mathord}{boldgreekuc}{"05}
\DeclareMathSymbol{\bfTheta}{\mathord}{boldgreekuc}{"02}
\DeclareMathSymbol{\bfRho}{\mathord}{boldgreekuc}{"52}
\DeclareMathSymbol{\bfSigma}{\mathord}{boldgreekuc}{"06}
\DeclareMathSymbol{\bfTau}{\mathord}{boldgreekuc}{"54}
\DeclareMathSymbol{\bfVartheta}{\mathord}{boldgreekuc}{"02} 
\DeclareMathSymbol{\bfOmega}{\mathord}{boldgreekuc}{"0A}
\DeclareMathSymbol{\bfVarphi}{\mathord}{boldgreekuc}{"08} 
\DeclareMathSymbol{\bfUpsilon}{\mathord}{boldgreekuc}{"07}
\DeclareMathSymbol{\bfZeta}{\mathord}{boldgreekuc}{"5A}

\DeclareFontFamily{U}{mp4sl}{}%
\DeclareFontShape{U}{mp4sl}{m}{n}{<->s * [\MPFourScale]cmmib10}{}
\DeclareSymbolFont{boldgreek}{U}{mp4sl}{m}{n}
\DeclareMathSymbol{\bfalpha}{\mathord}{boldgreek}{"0B}
\DeclareMathSymbol{\bfbeta}{\mathord}{boldgreek}{"0C}
\DeclareMathSymbol{\bfpsi}{\mathord}{boldgreek}{"20}
\DeclareMathSymbol{\bfdelta}{\mathord}{boldgreek}{"0E}
\DeclareMathSymbol{\bfepsilon}{\mathord}{boldgreek}{"0F}
\DeclareMathSymbol{\bfphi}{\mathord}{boldgreek}{"1E}
\DeclareMathSymbol{\bfgamma}{\mathord}{boldgreek}{"0D}
\DeclareMathSymbol{\bfeta}{\mathord}{boldgreek}{"11}
\DeclareMathSymbol{\bfiota}{\mathord}{boldgreek}{"13}
\DeclareMathSymbol{\bfxi}{\mathord}{boldgreek}{"18}
\DeclareMathSymbol{\bfkappa}{\mathord}{boldgreek}{"14}
\DeclareMathSymbol{\bflambda}{\mathord}{boldgreek}{"15}
\DeclareMathSymbol{\bfmu}{\mathord}{boldgreek}{"16}
\DeclareMathSymbol{\bfnu}{\mathord}{boldgreek}{"17}
\DeclareMathSymbol{\bfpi}{\mathord}{boldgreek}{"19}
\DeclareMathSymbol{\bfvartheta}{\mathord}{boldgreek}{"23}
\DeclareMathSymbol{\bfrho}{\mathord}{boldgreek}{"1A}
\DeclareMathSymbol{\bfsigma}{\mathord}{boldgreek}{"1B}
\DeclareMathSymbol{\bftau}{\mathord}{boldgreek}{"1C}
\DeclareMathSymbol{\bftheta}{\mathord}{boldgreek}{"12}
\DeclareMathSymbol{\bfomega}{\mathord}{boldgreek}{"21}
\DeclareMathSymbol{\bfvarphi}{\mathord}{boldgreek}{"27}
\DeclareMathSymbol{\bfchi}{\mathord}{boldgreek}{"1F}
\DeclareMathSymbol{\bfupsilon}{\mathord}{boldgreek}{"1D}
\DeclareMathSymbol{\bfzeta}{\mathord}{boldgreek}{"10}

\makeatother

\usepackage[dvips]{graphicx}

\usepackage{tikz}
\usetikzlibrary{bayesnet}
\usetikzlibrary{fit,positioning}

\renewcommand{\normalsize}{\fontsize{11}{16}\selectfont}
\renewcommand{\footnotesize}{\fontsize{10}{12}\selectfont}
\fontsize{11}{16}\selectfont

\newcommand{\widesim}[2][1.5]{
  \mathrel{\overset{#2}{\scalebox{#1}[1]{$\sim$}}}
}

\begin{document}
\bibliographystyle{apa}

\title{
A Doubly Latent Space Joint Model for Local Item and Person Dependence
in the Analysis of Item Response Data
}

\maketitle


\author{Ick Hoon Jin}
\affil{Department of Applied and Computational Mathematics and Statistics \break University of Notre Dame}

\author{Minjeong Jeon}
\affil{Graduate School of Education and Information Studies \break University of California, Los Angeles}

\reprints{Correspondence should be sent to\\

\noindent E-Mail: ijin@nd.edu\\
\noindent Phone:  574-631-2741\\
\noindent Fax:  \\
\noindent Website: \\}

\newpage\vspace*{24pt}

\RepeatTitle{
A Doubly Latent Space Joint Model for Local Item and Person Dependence
in the Analysis of Item Response Data
}

\begin{center}\vskip3pt

\vspace{32pt}

Abstract\vskip3pt

\end{center}

\begin{abstract}

Item response theory (IRT) models explain an observed item response as a function of a respondent's latent trait and the item's property.
IRT is one of the most widely utilized tools for item response analysis;
however,  local item and person independence, which is a critical assumption for IRT, is often violated in real testing situations.
In this article, we propose a new type of analytical approach for item response data that does not require standard local independence assumptions.
By adapting a latent space joint modeling approach, our proposed model
can estimate pairwise distances to represent the item and person dependence structures,
from which item and person clusters in latent spaces can be identified.
We provide an empirical data analysis to illustrate an application of the proposed method.
A simulation study was also provided to evaluate the performance of the proposed method in comparison to an existing method.

\end{abstract}

\vspace{\fill}\newpage

\section{Introduction}\label{sec:intro}

Item response theory (IRT) is a modeling framework commonly utilized within various academic disciplines
(including but not limited to psychology, education, political science, sociology, public health, and epidemiology) to analyze discrete data.
IRT models explain an observed response to a test item as a function of a respondent's latent trait
(unobserved continuous variable, such as cognitive ability) and item properties such as difficulty.

To illustrate IRT models, suppose we have a binary response $X_{ki}$ for person $k$ ($k=1,...,n$) to item $i$ ($i=1,...,p$).
The Rasch model \citep{rasch:60}, one of the most widely used IRT models, can be written as follows:
\begin{equation}\label{eq:Rasch}
P(X_{ki}=1 | \theta_k) = \frac{\exp(\theta_k + \beta_i)}{1 + \exp(\theta_k + \beta_i)},
\end{equation}
\noindent where $\theta_k$ is person $k$'s latent trait (or ability)
and $\beta_i$ is the easiness (or minus difficulty) of item $i$.
It is typically assumed that ability $\theta_k$ is a random effect
that is independently and identically distributed with $\theta_k \widesim{iid} N(0, \sigma_{\theta}^2)$.

The independence assumption for respondents is often violated due to person clustering (e.g, paired samples, nested samples).
To address such violation, researchers proposed including an additional random effect parameter
that can capture the dependence among respondents \citep[e.g.,][]{fox:01}.
However, this method cannot be applied when the person clustering structure is unknown,
for instance, when groups of students shared their answers during a test.

Another critical assumption to validate the use of IRT models is that the item responses are locally independent of one another
for a given value of the person's latent trait \citep{chen:97}, which is referred to as the local independence assumption \citep{mcdonald:82}.
When local independence holds, the joint probability of correct responses to an item pair ($i$, $j$ with $i \neq j$)
is the product of the probabilities of the two items:
\[ P(X_{ki} = 1, X_{kj}=1 |\theta_k)  = P(X_{ki} =1| \theta_k) \times P(X_{kj} =1|\theta_k).\]

Unfortunately, the local independence assumption is frequently violated during actual testing situations,
for example, when items are clustered based on their shared contents and stimulus (e.g., items within the same reading passage)
or wording (e.g., positively and negatively worded items).
In addition, nonignorable missingness can cause local dependence among items.
For instance, if a test taker fails to reach item $i$ in a speeded test, he or she will fail to reach items beyond item $i+1$, thus
creating local dependence among all omitted item responses \citep{chen:97}.

Although evaluating the presence of local item dependence is critical in IRT analysis,
detecting dependence is generally a challenging task \citep{liu:12}.
Researchers have proposed numerous test statistics for detecting local dependence
among item pairs or triplets \citep[e.g.,][]{chen:97, yen:84, glas:03},
but most of those statistics can only accommodate small tests that have a limited number of  items \citep{bishop:75}. 
Limited-information test statistics have been developed such as 
$M_r$ (with $r$ = 2, 3) statistics \citep{maydeu:05, maydeu:06}. 
These limited-information statistics utilize residuals
based on lower-order (e.g., first- and second-) margins of the entire contingency table of possible response patterns.
$M_r$ statistics can be used to test the overall fit of the model, but they do not pinpoint
the source of the misfit \citep{liu:12}.  
Polytomous item models or testlet models have also been used
to handle locally dependent items \citep[e.g.,][]{wainer:87, wilson:95};
however, those modeling approaches can only be utilized when locally dependent items are known a priori.


The aim of the current study is to propose a new item analytic method that can capture potential dependencies
among items and among respondents in item response data.
Our key idea is to expand a latent space modeling approach, which is typically used for social network data analysis,
for the purpose of analyzing binary item response data.
Our proposed approach detects the dependence structure of items from multi-layer person networks
as well as the dependence structure of people from multi-layer item networks.
The dependence structures on the item and person sides can be visually displayed in latent spaces.
Hence, the proposed method can effectively be used to identify item and person clustering.
Furthermore, our approach provides item and person parameter estimates that can be interpreted
as item and person parameters of a regular one-parameter IRT model or a Rasch model.

Our study contributes to the fields of psychometrics and statistics in several aspects:
first, we provide an item analysis strategy that can solve
item dependence and person dependence problems, without needing to know the dependence structures a priori.
Second, we provide a single step procedure that estimates item/person parameters, while allowing users to
examine item and person dependence structures simultaneously.
Note that to identify a dependence structure within item response data,
a two-step procedure is usually applied: fitting an IRT model (step 1) and then computing various test statistics (step 2).
Lastly, our work is a novel application of latent space modeling to item response data.
We extend an existing latent space model into a doubly latent space joint model
for the purpose of simultaneously analyzing item and person networks.

The remainder of this article is organized as follows.
First, we describe the proposed doubly latent space joint model for the analysis of item response data.
We then present the MCMC computational framework. 
In the application section, we illustrate an application of our proposed approach to analyze complex cognitive assessment data.
We then provide a simulation study to evaluate the performance of the
proposed method compared with an existing approach. 
We end our paper with a summary and some discussions.

\section{Doubly Latent Space Joint Model for Item Response Data}

\subsection{Latent Space Model}

Statistical approaches based on a latent Euclidean space have been a useful tool for analyzing dissimilarity data.
\citet{Oh:2001} first introduced the latent space concept to configure similarities/dissimilarities of objects in Bayesian multidimensional scaling.
Their work has been extended to model-based network data analysis, referred to as a latent space model (\citealp{Hoff:2002}; LSM).

Specifically, LSM introduces the distance between the latent position of ${\bf z}_k$
as a penalty in a logistic regression framework when considering interactions between nodes,
where (or actor or vertices) $k$ has an unknown position ${\bf z}_k$ in a $D$-dimensional Euclidean latent space.
The probability of a link between the pairs of nodes then depends on the distance between them.
Generally, the smaller the distance between two nodes in the latent space,
the greater the probability that they are connected to each other.

Let $N$ be the number of nodes in a network and ${\bf Y}$ be the $N \times N$ adjacency matrix\footnote{
An adjacency matrix is a square matrix to represent a network, whose elements indicate whether or not pairs of nodes are connected (creating edges) in the network.}
containing the network information, where $y_{kl} = 1$ if node $k$ and $l$ are connected and 0 otherwise.
The diagonal terms of the adjacency matrix are zero, $y_{kl} = 0$, unless node $k$ is self-connected.
Let ${\bf Z}$ be a $N \times D$ latent position matrix where each row ${\bf z}_k = (z_{k1}, \cdots, z_{kD})$ is the $D$-dimensional vector
indicating the position of node $k$ in the $D$-dimensional Euclidean space.
The LSM can then be written as
\begin{equation}\label{eq:LSM1}
P({\bf Y} \mid {\bf Z},  \beta) = \prod_{k \neq l}P(y_{kl} \mid {\bf z}_k, {\bf z}_l, \beta)
= \prod_{k \neq l} \frac{\exp\left(\beta  - ||{\bf z}_k - {\bf z}_l||\right)^{y_{kl}}}
{1 + \exp\left(\beta  - ||{\bf z}_k - {\bf z}_l||\right)},
\end{equation}
where $||{\bf z}_k - {\bf z}_l|| = \sqrt{\sum_{d=1}^D (z_{kd} - z_{ld})^2}$
is the Euclidean distance between nodes $k$ and $l$.
The  number of dimensions, $D$ (of the Euclidean space) is often selected as 2 or 3 for visual-display purposes.
To estimate the intercept $\beta$ and the latent positions ${\bf Z}$, a Bayesian approach is typically applied.
We assume that ${\bf z}_k$ are independent draws from a spherical multivariate normal distribution, so that
\[ {\bf z}_k \widesim{iid} \mbox{MVN}_d\left(0, \sigma_z^2 I_d\right). \]
We refer readers to \citealp{Hoff:2002}; \citealp{Handcock:2007}; \citealp{Krivitsky:2009}; \citealp{Raftery:2012}; \citealp{Rastelli:2015}
for additional references on latent space modeling.


The distances between Euclidean latent spaces are invariant under rotation, reflection, and translation (\citealp{Hoff:2002}; \citealp{Shortreed:2006}).
Thus, for each latent position matrix $Z$, there are an infinite number of possible positions that result in the same log-likelihoods.
This invariance property can cause a major problem in the parameter estimation of latent space models; because the model specifies the distances between actors,
the estimated latent position of actors may poorly represent the actual actor positions,
even though the distances between the actors may be accurately determined.
To correct for this non-identifiability problem,
we applied the post-processing of MCMC samples through Procrustes matching \citep{Borg:2005}.


\subsection{Latent Space Joint Model for Item Response Data}
\label{sec:LSM}

Suppose ${\bf X}_{n \times p}$ is a binary item response dataset
where $n$ is the number of respondents, $p$ is the number of items,
and $x_{ki}$ indicates a binary response to item $i$ for person $k$.
To apply a latent space model to item response data, we first need to construct
two sets of adjacency matrices, ${\bf Y}_{i, n \times n}$ for item $i$ and ${\bf U}_{k, p \times p}$ for person $k$
to represent the networks for items and for persons:
\begin{equation}\label{eq:construct}
 {\bf Y}_{i, n \times n} = \big\{y_{i,kl}\big\} = \big\{ x_{ki} x_{li} \big\}
 \qquad \mbox{and} \qquad
 {\bf U}_{k, p \times p} = \big\{u_{k,ij}\big\} = \big\{ x_{ki} x_{kj} \big\},
\end{equation}
\noindent
where ${\bf Y}_{i, n \times n}$ and ${\bf U}_{k, p \times p}$ are undirected networks.
Specifically,  $y_{i,kl}$ takes 1 if persons $k$ and $l$ give a correct answer to item $i$ ($i=1,...,p$) and 0 otherwise;
that is, $x_{ki} x_{li}$ indicates an interaction, which can be seen as dependence between person $k$ and person $l$ for item $i$.
Similarly, $u_{k,ij} =1$ if person $k$ ($k=1,...,n$) gives correct answers to items $i$ and $j$ ($i \neq j$) and 0 otherwise
with $ x_{ki} x_{kj} $ indicating an interaction (or dependence) between items $i$ and $j$ for person $k$.
It is important to note that we allow for dependence between pairs of items within person
as well as dependence between pairs of respondents for an item.
This idea makes our approach unique compared to typical IRT models and the recently developed,
Ising model-based approaches  \citep{vanBorkulo:2014, Kuris:2016} that require local independence assumptions.


%

\begin{figure}[hbtp]
\begin{center}
\begin{tabular}{cc}
(a) LSJM for Items &  (b) LSJM for Persons\\
\includegraphics[width=0.45\textwidth]{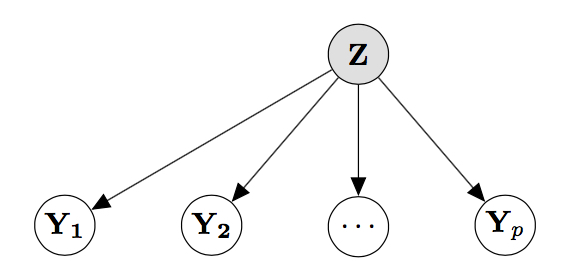} &
\includegraphics[width=0.45\textwidth]{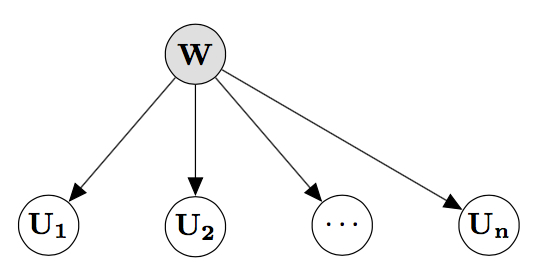} \\
\end{tabular}
\end{center}
\caption{\label{fig:LSM-IRT}
Latent Space Joint Model for Items (a) and Latent Space Joint Model for Persons (b).
${\bf Z} = \{{\bf z}_k\}$ and ${\bf W} = \{{\bf w}_i\}$
denotes a latent space for $Y_i$ and $U_k$, respectively,
where $i = 1, \cdots, p$ and $k = 1, \cdots, n$. The latent space joint model for items
contains $n$ person latent spaces  and the latent space joint model for persons includes $p$ item latent spaces.
}
\end{figure}

Figure \ref{fig:LSM-IRT} illustrates LSJM for items and persons.
In Figure \ref{fig:LSM-IRT}, ${\bf Z} = \left\{{\bf z}_k\right\}$ and ${\bf W} = \left\{{\bf w}_i\right\}$ denote
latent spaces for person network ${\bf Y}_i$ (for item $i$) and item network $U_k$ (for person $k$),  respectively.
Note that adjacency matrices ${\bf Y}_i$ and ${\bf U}_k$ are defined for each item ($i$) and for each person ($k$).
Hence, there are multiple networks on the item and person sides (multiple networks with the same set of nodes
(e.g., items or persons in the current paper) are called multi-layer or multiplex networks).
It is assumed that multi-layer (or multiplex) networks ${\bf Y}$ are assumed to be conditionally independent given the latent space ${\bf Z}$
where ${\bf z}_k \sim N\left(0, \sigma_z^2 {\bf I}_D\right)$
is determined in a $D$-dimensional latent space.
Similarly, multi-layer networks ${\bf U}$ are conditionally independent
given the latent space ${\bf W}$ where ${\bf w}_i$ $\sim$ $N(0, \sigma_w^2 {\bf I}_D)$
is determined in a $D$-dimensional latent space.

Latent space ${\bf Z}$ summarize the latent feature information of the respondents from multi-layer networks ${\bf Y}$,
while latent space ${\bf W}$ summarize the latent feature
information of the items from multi-layer networks ${\bf U}$.
\vskip -0.75cm
\begin{center}
[Insert Figure \ref{fig:LSM-IRT}]
\end{center}
\vskip -0.25cm
We first specify LSJM for items illustrated in Figure \ref{fig:LSM-IRT}(a) as follows:
\begin{equation}\label{eq:LSM-ITEM}
P\Big({\bf Y} \mid {\bf Z}, \boldsymbol\beta \Big)
= \prod_{i=1}^p P\left({\bf Y}_i \mid {\bf Z}; \beta_i\right)
= \prod_{i=1}^p \prod_{k \neq l} \frac{\exp\left(\beta_i - ||{\bf z}_k - {\bf z}_l|| \right)^{y_{i,kl}}}
{1 + \exp\left(\beta_i - ||{\bf z}_k - {\bf z}_l||\right)},
\end{equation}
where $\beta_i$ is the intercept parameter for item $i$, ${\bf z}_k$ and ${\bf z}_l$ indicate the latent positions for person $k$ and person $l$.
Here $\beta_i$ can be interpreted as the (inverse logit transformed) probability of correctly answering item $i$
when respondents $k$ and $l$ have the same latent space positions (in other words, when respondents $k$ and $l$ have the same ability levels).
Note that $\beta_i$ is conceptually similar to the item easiness parameter of the Rasch model.
The key difference is that $\beta_i$ in LSJM is determined by
whether pairs of respondents, with similar or different abilities, jointly answer the item correctly.
For instance, a large $\beta_i$ is obtained when pairs of respondents with highly different abilities
(or with a large distance in their latent space positions) tend to answer the item correctly.
On the other hand, a small  $\beta_i$ is obtained when pairs of respondents fail to correctly answer the item together.
In this sense, one can utilize the item intercept parameter estimates to discuss and compare
the overall easiness levels of individual items.

The prior distributions for the model parameters are specified
as $p(\beta_i) \sim N\left(0, \sigma_\beta^2\right)$, $p({\bf z}_k \mid \sigma_z^2) \sim N\left(0, \sigma_z^2 I_D\right)$,
and $p(\sigma_z^2) \sim \mbox{Inv-Gamma}\left(a_{\sigma_z}, b_{\sigma_z}\right)$ with fixed $\sigma_\beta^2$, $a_{\sigma_z}$, and $b_{\sigma_z}$.
Since $\sigma_z^2$ quantifies the contribution of latent spaces, we assign a hyper prior to estimate $\sigma_z^2$.
Then, the conditional posterior distribution for $\beta_i$, ${\bf z}_k$, and $\sigma_z^2$ can be specified as
\begin{equation}\label{eq:post-lsm-item}
\begin{split}
\pi\left(\beta_i \mid {\bf Y}_i, {\bf Z}\right)
&\propto \pi(\beta_i) \prod_{k \neq l} \frac{\exp\left(\beta_i - ||{\bf z}_k - {\bf z}_l||\right)^{y_{i,kl}}}
{1 + \exp\left(\beta_i - ||{\bf z}_k - {\bf z}_l||\right)}\\
\pi\left({\bf z}_k \mid {\bf Y}, \boldsymbol\beta, \sigma_z^2 \right)
&\propto \pi\left({\bf z}_k \mid \sigma_z^2\right) \prod_{i=1}^p p\left({\bf Y}_i \mid {\bf z}_k, \beta_i\right),\\
\pi\left(\sigma_z^2 \mid {\bf Z}\right) &\propto \pi(\sigma_z^2) \prod_{k=1}^n p\left({\bf z}_k \mid \sigma_z^2\right).
\end{split}
\end{equation}


Next, we specify LSJM for persons illustrated in Figure \ref{fig:LSM-IRT}(b) as follows:
\begin{equation}\label{eq:LSM-Individual}
P\Big({\bf U} \mid {\bf W}, \boldsymbol\theta \Big)
= \prod_{k=1}^n P\left({\bf U}_k \mid {\bf W}; \theta_k\right)
= \prod_{k=1}^n \prod_{i \neq j} \frac{\exp\left(\theta_k - ||{\bf w}_i - {\bf w}_j||\right)^{u_{k,ij}}}
{1 + \exp\left(\theta_k - ||{\bf w}_i - {\bf w}_j||\right)},
\end{equation}
where $\theta_k$ is the intercept parameter for person $k$, ${\bf w}_i$ and ${\bf w}_j$ indicate the latent positions for item $i$ and item $j$, respectively.
Here $\theta_k$ can be interpreted as the (inverse logit transformed) probability of correctly answering items $i$ and $j$ for person $k$
when items $i$ and $j$ have the same latent space positions
(in other words, if person $k$ gives a correct answer to item $i$, then he/she has also given a correct answer to item $j$).
Note that $\theta_k$ is conceptually similar to the person ability parameter of the Rasch model,
although a key difference is that $\theta_k$ in LSJM is determined
by whether a person correctly answers pairs of items with similar or different levels of easiness.
For example, a large $\theta_k$ is obtained when the respondent tends to answer pairs of items with highly different levels of easiness
(or items with a large distance in their latent positions).
A small  $\theta_k$ is obtained when the person fails to  answer many pairs of items correctly.
Thus, one can use the person intercept parameter estimates
to compare the level of abilities (or latent traits) among the respondents.

The prior distributions are specified for the parameters
as $p(\theta_k) \sim N\left(0, \sigma_\theta^2\right)$, $p({\bf w}_i \mid \sigma_w^2) \sim N\left(0, \sigma_w^2 I_D\right)$,
and $p(\sigma_w^2) \sim \mbox{Inv-Gamma}\left(a_{\sigma_w}, b_{\sigma_w}\right)$ with fixed $\sigma_{\theta}^2$, $a_{\sigma_w}$, and $b_{\sigma_w}$.
Since $\sigma_w^2$ quantifies the contribution of latent spaces, we assign a hyper prior to estimate $\sigma_w^2$.
The conditional posterior distribution for $\theta_k$,  ${\bf w}_i$, and $\sigma_w^2$
can then be specified as
\begin{equation}\label{eq:post-lsm-individual}
\begin{split}
\pi\left(\theta_k \mid {\bf U}_k, {\bf W}\right)
&\propto \pi(\theta_k) \prod_{i \neq j} \frac{\exp\left(\theta_k - ||{\bf w}_i - {\bf w}_j||\right)^{u_{k,ij}}}
{1 + \exp\left(\theta_k - ||{\bf w}_i - {\bf w}_j||\right)}.\\
\pi\left({\bf w}_i \mid {\bf U}, \boldsymbol\theta, \sigma_w^2\right)
&\propto \pi({\bf w}_i \mid \sigma_w^2) \prod_{k=1}^n p\left({\bf U}_k \mid {\bf w}_i, \theta_k\right),\\
\pi\left(\sigma_w^2 \mid {\bf W}\right) &\propto \pi(\sigma_w^2) \prod_{i=1}^p p\left({\bf w}_i \mid \sigma_w^2\right).
\end{split}
\end{equation}


\subsection{Doubly Latent Space Joint Model for Item Response Data}
\label{sec:LSJM}

\begin{figure}[htbp]
\begin{center}
\includegraphics[width=0.9\textwidth]{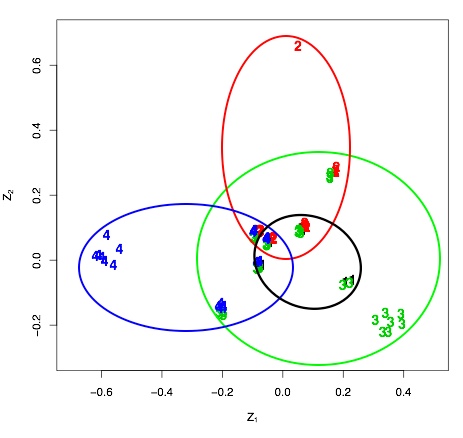}
\end{center}
\caption{\label{fig:link}
A person latent space constructed for an example dataset (with four items and 60 respondents).
The latent positions displayed in the figure are obtained from the MCMC procedure.
The locations of numbers indicate the latent positions of individual respondents.
Each number represents the item number that the respondents in the position
correctly answered. For instance, the number 4s (in blue) on the far left side
indicate the positions of the respondents who correctly answered Item 4 only.
Respondents around the center have overlapping numbers (1,2,3, and 4)
since they gave correct answers to all four items.
}
\end{figure}

For simultaneous estimation, we must combine the two LSJM models for items and persons, constructed in the Section of
Latent Space Joint Model for Item Response Data.
Unfortunately, the two models cannot be directly integrated
because of the dimensional mismatch in ${\bf Z}_{n \times n}$ and ${\bf W}_{p \times p}$.
To resolve this issue,
we assume that an item latent space can be computed based on a person latent space.
This assumption is motivated by the fact that
(1) the two latent spaces (for items and persons) are essentially determined based on a single item response dataset, and
(2) the latent space of a respondent is essentially determined based on his/her item response patterns.
To further elucidate the reasoning behind our assumption, we provide an illustration in Figure \ref{fig:link}
that displays a person latent space for an example dataset that includes four items and 60 respondents.

\begin{center}
[Insert Figure \ref{fig:link}]
\end{center}

In Figure \ref{fig:link}, each respondent is represented by the item number that he/she answered correctly.
For example, those respondents who gave correct answers to all items (1, 2, 3, 4) are placed around the origin (0,0) of the plot.
Those who gave correct answers to item 1 and 3 are located right-below of the first group (0.2, -0.01).
This way, we can identify respondents who gave correct answers to individual items.
The figure shows four respondent groups (who gave correct answers to items 1, 2, 3, and 4, respectively) with four colored ellipses:
(1) the black ellipse indicates the respondents who correctly answered item 1 correctly,
(2) the red ellipse denotes those who answered item 2 correctly,
(3) the green ellipse indicates those who answered item 3  correctly,
and (4) the blue ellipse indicates those who gave a correct answer to item 4.\footnote{
The black ellipse is completely overlapped with the green ellipse, indicating that there is a strong dependence between items 1 and 3.}
Note that the center of each ellipse (that represents a respondent group) can be seen as the position of each item in the latent space.
This means that an item latent position is actually the average of the respondent latent positions who correctly answered the corresponding item.


Based on this reasoning, we define the latent space of item $i$ (${\bf w}_i$)
as a function of the latent spaces of all respondents (${\bf Z}$), where the function is defined as follows:
\begin{equation}
{\bf w}_i =  f_i({\bf Z})  = \sum_{k=1}^n \frac{x_{ki}{\bf z}_{k.}}{\sum_{k=1}^n x_{ki}}.
\end{equation}
\noindent That is, ${\bf w}_i$ is regarded as an average of latent space collections for the respondents who give a correct answer to item $i$.

\begin{figure}[htbp]
\begin{center}
Doubly Latent Space Joint Model for Item Response Data
\vskip .3cm
\includegraphics[width=0.9\textwidth]{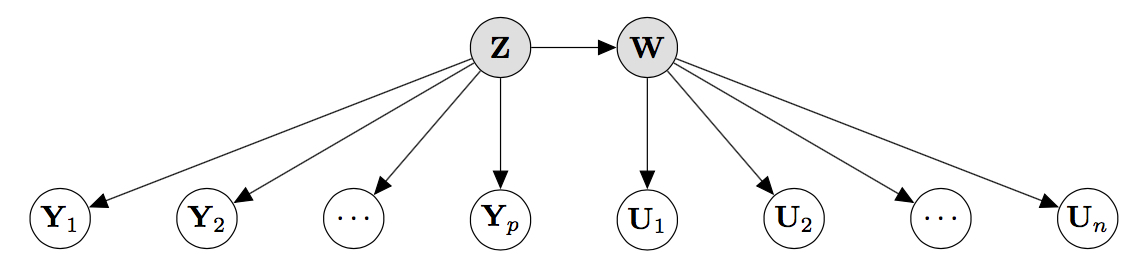}
\end{center}
\caption{\label{fig:LSJM-IRT}
Doubly Latent Space Joint Model.
${\bf Z} = \{{\bf z}_k\}$ and ${\bf W} = \{{\bf w}_i\} = \{f_i({\bf Z})\}$
denotes a latent space of item property matrix $Y_i$ and personal characteristics matrix $U_k$, respectively,
where $i = 1, \cdots, p$ and $k = 1, \cdots, n$.
}
\end{figure}

Based on this assumption, the two LSJM models for items and persons
can be integrated and jointly estimated.
We refer to the resulting, integrated model as
a doubly latent space joint model (DLSJM) for item response data.
Figure \ref{fig:LSJM-IRT} illustrates DLSJM.

\begin{center}
[Insert Figure \ref{fig:LSJM-IRT}]
\end{center}

DLSJM assumes that two sets of multi-layer networks
${\bf Y}$ and ${\bf U}$ are conditionally independent given latent space ${\bf Z}$.
Note that since latent spaces are defined based on pairwise distances among items and among persons, assuming conditional independence given the item and person latent spaces is different from the typical local independence assumptions adopted in standard IRT analysis.
Mathematically, DLSJM can be expressed as
\begin{equation}\label{eq:LSJM}
\begin{split}
P\Big(&{\bf Y}, {\bf U} \mid {\bf Z}, \boldsymbol\beta, \boldsymbol\theta\Big)
= \prod_{i=1}^p P\big({\bf Y}_i \mid {\bf Z}, \beta_i\big) \prod_{k=1}^n P\big({\bf U}_k \mid {\bf Z}, \theta_k\big)\\
&= \prod_{i=1}^p \prod_{k \neq l} \frac{\exp\left(\beta_i - ||{\bf z}_k - {\bf z}_l||\right)^{y_{i,kl}}}
{1 + \exp\left(\beta_i - ||{\bf z}_k - {\bf z}_l||\right)}
\prod_{k=1}^n \prod_{i \neq j} \frac{\exp\left(\theta_k - ||f_i({\bf z}) - f_j({\bf z})||\right)^{u_{k,ij}}}
{1 + \exp\left(\theta_k - ||f_i({\bf z}) - f_j({\bf z})||\right)}.
\end{split}
\end{equation}
Here the interpretations of item intercept parameter $\beta_i$ and person intercept parameter $\theta_k$
remain the same as in Equations (\ref{eq:LSM-ITEM}) and (\ref{eq:LSM-Individual}).
With the prior distributions,
$p(\beta_i) \sim N(0, \sigma_\beta^2)$, $p(\theta_k) \sim N(0, \sigma_\theta^2)$,
$p({\bf z}_k \mid \sigma_z^2) \sim N\left(0, \sigma_z^2 {\bf I}_D\right)$, and $p(\sigma_z^2) \sim \mbox{Inv-Gamma}\left(a_{\sigma}, b_{\sigma}\right)$
with fixed $\sigma_\beta^2$, $\sigma_\theta^2$, $a_{\sigma}$ and $b_{\sigma}$,
the conditional posterior distribution for $\beta_i$, $\theta_k$, ${\bf z}_k$, and $\sigma_z^2$ can be specified as follows:
\begin{equation}\label{eq:post-lsjm}
\begin{split}
\pi\left(\beta_i \mid {\bf Y}_i, {\bf Z}\right)
&\propto \pi(\beta_i) \prod_{k \neq l} \frac{\exp\left(\beta_i - ||{\bf z}_k - {\bf z}_l||\right)^{y_{i,kl}}}
{1 + \exp\left(\beta_i - ||{\bf z}_k - {\bf z}_l||\right)},\\
\pi\left(\theta_k \mid {\bf U}_k, {\bf Z}\right)
&\propto \pi(\theta_k) \prod_{i \neq j} \frac{\exp\left(\theta_k - ||f_i({\bf z}) - f_j({\bf z})||\right)^{u_{k,ij}}}
{1 + \exp\left(\theta_k - ||f_i({\bf z}) - f_j({\bf z})||\right)},\\
\pi\Big({\bf z}_k \mid {\bf Y}, {\bf U}, \boldsymbol\beta, \boldsymbol\theta\Big)
&\propto \pi({\bf z}_k \mid \sigma_z^2) \prod_{i=1}^p P\left({\bf Y}_i \mid {\bf z}_k, \beta_i\right)
\prod_{k=1}^n P\left({\bf U}_k \mid f_i({\bf z}_k), \theta_k\right),\\
\pi\left(\sigma_z^2 \mid {\bf Z}\right) &\propto \pi(\sigma_z^2) \prod_{k=1}^n p\left({\bf z}_k \mid \sigma_z^2\right).
\end{split}
\end{equation}

\subsection{Comparisons with Existing Approaches}

A variety of methods have been proposed in psychometrics to explore item response data
and to identify item clustering (or dimensions) or person clustering (or latent classes) structures.
In this section, we discuss some of existing methods that are comparable
to our DLSJM approach.

First, the DLSJM approach may seems similar to multidimensional scaling (MDS).
MDS analyzes proximity (or similarity) data represented by spatial distance models (where space often refers to Euclidean space as in the DLSJM).
Similar to DLSJM, MDS represents a set of stimuli (e.g., items) as points
in a multidimensional Euclidean space in such a way that those points for similar stimuli
are located close together, while those for dissimilar stimuli are located far apart \citep{takene:07}.
\citet{delay:91} reported that MDS is a useful method for assessing dimensionality of item response data
when Euclidean distances were used as item proximity measures.
The main difference between MDS and DLSJM, however, lies in the that
MDS is a non-modeling method that has usually been applied to identifying dimensions (or clusters) of items only,
while DLSJM is a model-based method that is designed to identify both
item and person clusters in item response data.

Second, the DETECT procedure \citep{stout:96} 
may also be seen as being  comparable to the DLSJM approach.
The DETECT method, however, is a non-model-based (or nonparametric) method
that is designed to detect dimensions (or clusters) of item response data based on item-pair covariance.
Specifically, the DETECT procedure finds the number of dimensions of data by searching through all possible item partitions
until the optimal partition is found. Unlike the DETECT method,
the DLSJM approach is a model-based method that aims to capture both item and person clusters,
while providing item and person parameters similar to a regular IRT model.

Third, one may also view an exploratory IRT approach to be similar to the DLSJM method
in that exploratory IRT analysis is applied to identify dimensions (or clustering) of item response data as in DLSJM.
Although the use of the two approaches may be similar in terms of item cluster identification,
there is a technical difference between the two methods:
the exploratory IRT analysis extracts factors (or dimensions) based on the amount of variance explained \citep{Kamata:08},
while DLSJM is based on pairwise distances among items to identify item clusters.
Thus, the factor structure (based on an item loading matrix) obtained from an exploratory IRT
is not necessarily equivalent to the item dependence (or clustering) structure (based on a pairwise distance matrix). 

Lastly, the DLSJM approach may also be seen similar to a finite mixture approach, which is known as a model-based clustering method \citep{Handcock:2007}.
In psychometrics, the mixture IRT approach \citep[e.g.,][]{rost:90}
has been employed to identify clusters of individual respondents.
An important technical difference between the two approaches is that
mixture IRT analysis requires respondents' independence within a cluster (or latent class),  
whereas the DLSJM approach does not require such a within-cluster independence assumption.
In addition, mixture IRT analysis still require the local item independence assumption,
the independence of item responses within a person (or conditional on a person's ability) \citep{Gollini:2016}.

In summary, even though several existing psychometric approaches are available to identify clusters of item or respondents,
they are often non-modeling methods and/or are unable to identify both person and item clusters in item response data.
Our proposed DLSJM approach is a unique contribution to psychometrics, because of its ability to identify both person and item clustering structures
simultaneously, while providing item and person parameters 
that capture heterogeneity among items as well as among respondents.

%

\section{Markov Chain Monte Carlo Estimation}
\label{sec:MCMC}

To estimate the DLSJM's model parameters $\boldsymbol\beta$, $\boldsymbol\theta$, and latent positions ${\bf Z}_k$,
we apply a standard Bayesian approach with the Metropolis-Hasting algorithm
(\citealp{Hoff:2002}; \citealp{Handcock:2007}; \citealp{Krivitsky:2009}; \citealp{Raftery:2012}; \citealp{Rastelli:2015}).
One iteration of the  Markov chain Monte Carlo (MCMC) sampler for DLSJM can be described as follows:
\begin{enumerate}
\item For each $k$ in a random order, propose a value ${\bf z}_k'$ from the proposal distribution $\varphi_{1k}(\cdot)$ and accept with probability
\[
r_z\Big(z_k', z_k^{(t)}\Big) = \frac{\pi\Big({\bf z}_k' \mid {\bf z}_{-k}, {\bf Y}, {\bf U}, \boldsymbol\beta, \boldsymbol\theta\Big)}
 {\pi\Big({\bf z}_k^{(t)} \mid {\bf z}_{-k}, {\bf Y}, {\bf U}, \boldsymbol\beta, \boldsymbol\theta\Big)}
 \frac{\varphi_{1k}({\bf z}_k' \to {\bf z}_k^{(t)})}{\varphi_{1k}({\bf z}_k^{(t)} \to {\bf z}_k')},
\]
where ${\bf z}_{-k}$ are all components of ${\bf Z}$ except ${\bf z}_{k}$.
\item Update $\sigma_z^2$ using inverse-gamma distribution.
\item Propose $\beta_i'$ from the proposal distribution $\varphi_2(\cdot)$ and accept with probability
\[
r_{\beta}\Big(\beta_i', \beta_i^{(t)}\Big) = \frac{\pi(\beta_i' \mid {\bf Y}_i, {\bf Z})}{\pi(\beta_i^{(t)} \mid {\bf Y}_i, {\bf Z})}
\frac{\varphi_2(\beta_i' \to \beta_i^{(t)})}{\varphi_2(\beta_i^{(t)} \to \beta_i')}
\]
\item Propose $\theta_k'$ from the proposal distribution $\varphi_3(\cdot)$ and accept with probability
\[
r_{\theta}\Big(\theta_k', \theta_k^{(t)}\Big) = \frac{\pi(\theta_k' \mid {\bf U}_k, {\bf Z})}{\pi(\theta_k^{(t)} \mid {\bf U}_k, {\bf Z})}
\frac{\varphi_3(\theta_k' \to \theta_k^{(t)})}{\varphi_3(\theta_k^{(t)} \to \theta_k')}.
\]
\end{enumerate}

Running the MCMC sampler for DLSJM is computationally demanding (especially for large datasets),
because (1) updating ${\bf Z}$ requires calculating  $n \times (n - 1) \times (p - 1)$ terms of the log-likelihood, and
(2) updating of $\boldsymbol\beta$ and $\boldsymbol\theta$ requires
calculating  all $p \times {n \choose 2}$ and $n \times {p \choose 2}$ terms of the log-likelihood \citep{Raftery:2012}.
Both updates need at least $O(n^2 p)$ calculations at each iteration of the MCMC algorithm.
That is, the computational cost of DLSJM becomes quickly exorbitant 
as the number of respondents and the number of items increase.

To alleviate the computational burden of DLSJM's MCMC sampler,
we utilize a parallel computing technique ({\tt OpenMP}).
Alternatively, one may consider using techniques that can reduce the computational complexity,
e.g., by approximating the log-likelihood with a case-control approximate likelihood \citep{Raftery:2012}
or by estimating the parameters based on the variational approximation with EM algorithm \citep{Gollini:2016}.


To apply the described MCMC algorithm to estimate DLSJM, we need to determine the proposal distribution $\varphi_1(\cdot)$ for ${\bf z}_k$.
Since the edges impose constraints on the positions of the nodes,
the latent positions for high-degree nodes do not tend to move freely; as the result,
those latent positions concentrate near their center, creating potential mixing problems.
On the other hand, the latent positions for low-degree nodes tend to move freely and are located in the perimeters.

Therefore, for an efficient mixing of the MCMC chain, we adjusted the variance of the proposal distributions based on the degree of a node.
For instance, we applied a small jumping rule for heavily-connected nodes
and a large jumping rule for lightly-connected nodes so that the ideal acceptance rates (20\% to 40\%) could be achieved.

To correct for potential rotation or translation problems that might have occurred during the MCMC sampling process \citep{Friel:2016},
we post-processed the MCMC samples using Procrustes matching \citep{Borg:2005}.
To implement a Procrustes matching method, we followed the procedure used by \citet{Friel:2016} which can be summarized as follows:
\begin{enumerate}
\item[1.] To find a reference set of latent positions, we picked out the latent positions in MCMC samples
that achieved the highest value of the full log posterior density.
\item[2.] We then applied Procrustes matching to each of the MCMC samples, using the reference set of latent positions.
\end{enumerate}

To check the convergence of the MCMC algorithm,
we utilize the distance measures between the pairs of respondents' and items' latent positions
(due to the latent space invariance property).
The convergence of the distance measures for DLSJM
is guaranteed, regardless of the fact that item latent spaces are a function of the respondent latent spaces,
because the distance measures are included in the MCMC acceptance ratio.
Trace plots in Section B in the supplement materials confirm that the distance measures for item latent spaces
have good convergence in our DRV data analysis.

\section{Application}\label{sec:drv}

Here we illustrate an application of the DLSJM to an empirical dataset.
We first describe the data utilized in this study
and provide the DLSJM analysis results in detail.

\subsection{Data}

We used an item response dataset based on the Competence Profile Test of Deductive Reasoning - Verbal assessment \citep[DRV;][]{spiel:01, spiel:08}.
Deductive reasoning is a logical process in which a conclusion is reached based on the agreement of multiple premises that are assumed to be true.
The DRV assessment was developed based on Piaget's cognitive developmental theory \citep{piaget:71} for evaluating children's cognitive developmental stages.
According to Piaget's theory, children move through four qualitatively different cognitive developmental stages:
the sensorimotor, the preoperational, the concrete-operational, and the formal-operational stages.
The progress from one stage to another requires children to apply a major reorganization of their thought process \citep{draney:07b}.
For instance, children in the concrete-operational stage are expected to perform logical operations only on concrete objects.
Children in the formal-operational stage are expected to perform logical operations on abstractions as well as concrete objects.
The DRV test focuses on identifying which of the two developmental stages the children are in.


The DRV test was constructed based on several difficulty factors with systematic variations \citep{spiel:01}.
The first factor is the Type of inference. This factor concerns four inference types
based on premises and conclusions:
(1) Modus Ponens (MP; A, therefore B), (2) Negation of Antecedent (NA; Not A, therefore B or not B),
(3) Affirmation of Consequent (AC; B, therefore A or not A), and (4) Modus Tollens (MT; Not B, therefore not A).
Modus Ponens (MP) and Modus Tollens (MT) involve bi-conditional conclusions (with ``yes'' or ``no'' response options),
while negation of antecedent (NA) and affirmation of consequent (AC) also include a ``perhaps'' option.
The NA and AC items are also called logical fallacy items because they provoke a logically incorrect conclusion.
For example, an AC item is given by ``Tom is lying in his bed. Is Tom
ill'' (the correct answer is ``perhaps'').
The second design factor is Content of the conditional.
This factor includes three content types: (a) Concrete (CO), (ab) Abstract (AB), and (c)
Counterfactual (CF). An example of a CF item is ``If an object is put into boiling water, it
becomes cold''. The third design factor is Presentation of the antecedent. The antecedent can
be presented with negation (NE) or without negation (NN). For example, with negation, an
item can be given as ``If the sun does not shine, Peter wears blue pants''.

Researchers found that children at the concrete-operational stage tend to treat all four inferences as bi-conditional,
thereby giving incorrect responses to logical fallacy items \citep[e.g.,][]{evans:93,  janveau:99}.
As cognitive development progresses, the performance on fallacy items (NA, AC) usually improves.
However, the performance on bi-conditional items (MT and MP) may worsen because
children who notice the uncertainty of the fallacies often overgeneralize \citep[e.g.,][]{byrnes:86, markovits:98}.
In addition, it has been reported that concrete items are usually easier than abstract and counterfactual items,
while items with negations are more difficult than items without negations \citep[e.g.,][]{roberge:78}.
Difficulty differences between abstract and counterfactual items seem to be unclear \citep[e.g.,][]{overton:85}.

\subsection{Previous Analytic Approaches and Issues}

Prior studies that analyzed the DRV data applied finite-mixture IRT models \citep{spiel:01, draney:07}
for identifying students' cluster memberships (that represent their developmental stages).
For instance, \citet{spiel:01} applied a mixture Rasch model and
identified three latent classes of respondents, while describing Class 1 as
participants who correctly solved only MP and MT items (hence, the
concrete-operational stage), Class 2 as those who performed better in NA and AC item than
Class 1 students (hence, the formal-operational stage),
and Class 3 as those who showed mixed performance and therefore were in the transition
between the concrete-operational stage and the formal-operational stage.
These studies have confirmed that there are heterogeneous respondent grouping in the DRV test
data. However, they still did not consider the possibility of potential item groupings (or dependence) among the DRV test items
(although such local item independence assumption is likely to be violated in the data due to items' shared design factors).
Hence, we applied the DLSJM approach to the DRV data
so that we could simultaneously examine both person and item groupings of the data. 



\subsection{Analysis and Results}

The DRV assessment data collected by \citet{spiel:01} were used for data analysis.
The DRV test, consisting of 24 items based on three design (difficulty) factors described earlier, was administered to 418 secondary school
students (162 females and 256 males) in Graz, Austria. 
There was approximately the same
number of students in grades 7 through 12 (age 11 through 18). The students' responses were coded dichotomously, with 1 for correct, and 0 for incorrect responses.

Note that although Piaget theorized that the transition from concrete-operational stage to
formal-operational stage occur around age 12, numerous empirical studies reported that the
transition might not happen even in late adolescence and early adulthood. For that reason,
 \citet{spiel:01} targeted the school-aged-participants (in the age range of 11 to 18) for their
investigation.

The DLSJM was applied to analyze the DRV item response data.
MCMC was implemented as described in the Section of Markov Chain Monte Carlo Estimation. 
The MCMC run consisted of 55,000 iterations with
the first 5,000 iterations being discarded as a burn-in process.
From the remaining 50,000 iterations, 5,000 samples were collected at a time space of 10 iterations.
A two-dimensional Euclidean space was used for item and person latent spaces.\footnote{
Note that the `dimensions' of a latent space are different from `dimensions' in multidimensional IRT.
The dimensions of a latent space are arbitrary coordinates to define a Euclidean space for pairwise distance among items as well as among persons.
Clusters of items in the latent space, which will be detected as the result of the DLSJM estimation, may be seen as the dimensions (or factors) of items.}
A jumping rule was set to 0.1 for $\varphi_2(\cdot)$ and to 3.0 for $\varphi_3(\cdot)$.
Different jumping rules were applied for $\varphi_1(\cdot)$
(a detailed description of the jumping rules were provided in the supplementary material).
In addition, we fixed $\sigma_{\theta} = \sigma_{\beta} = 10.0$ and $a_\sigma = b_\sigma$ = 0.01.

\subsubsection{Person Dependence Structure}

To summarize the estimates of the person intercept parameter ($\theta$),
we grouped all test participants based on their total scores;
then, we calculated a five number summary of the $\theta$ estimates per group,
which is provided in Section A.2 of the supplementary material.
As expected, the $\theta$ estimates tend to increase as more items are correctly answered (or the total score increases).
Note that there seems to be little difference in the $\theta$ estimates between for cases with the total scores of 0 and 1.
This is because when the total score is 1 or 0, the resulting adjacency (or network) matrix is an empty matrix. 

To aid in the visualization of the person dependence structure,
we applied a spectral clustering technique \citep{Ng:2002, Luxburg:07} to the pairwise distance measures of persons,
which utilizes the spectrum (eigenvalues) of a similarity matrix for clustering.
The spectral clustering was implemented with the {\tt specClust} function in the {\tt kknn} R package \citep{Hechenbichler:2004}.
For the analysis, we used the negative exponential of the estimated distance matrices
between the pairs of person's post-processed latent spaces as the similarity matrices.
We utilized $k$-nearest neighbor graphs, while selecting  $k$ such that the explanatory power of clustering was maximized.
Specifically, $k= \lfloor .50 \times n \rfloor$ was chosen with approximately 87.6\% of the variance explained (where $n$ is the total number of persons).
As in the prior study \citep{spiel:01}, 
we specified three clusters that would correspond to the concrete-operational stage, the formal-operational stage, and the transition between the stages.

\begin{figure}[htbp]
\centering
\begin{tabular}{cc}
(a) & (b) \\
\includegraphics[width=0.5\textwidth]{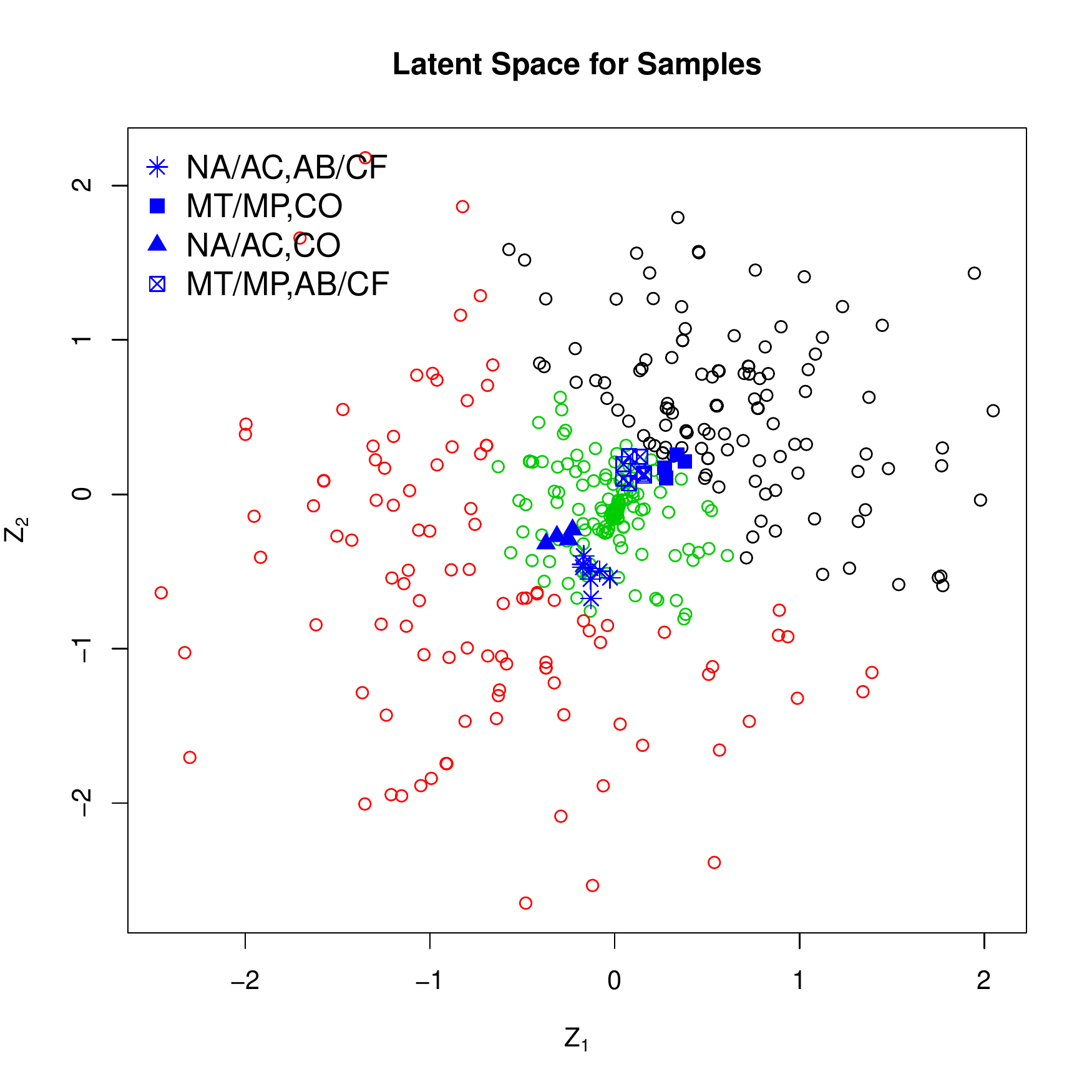} &
\includegraphics[width=0.5\textwidth]{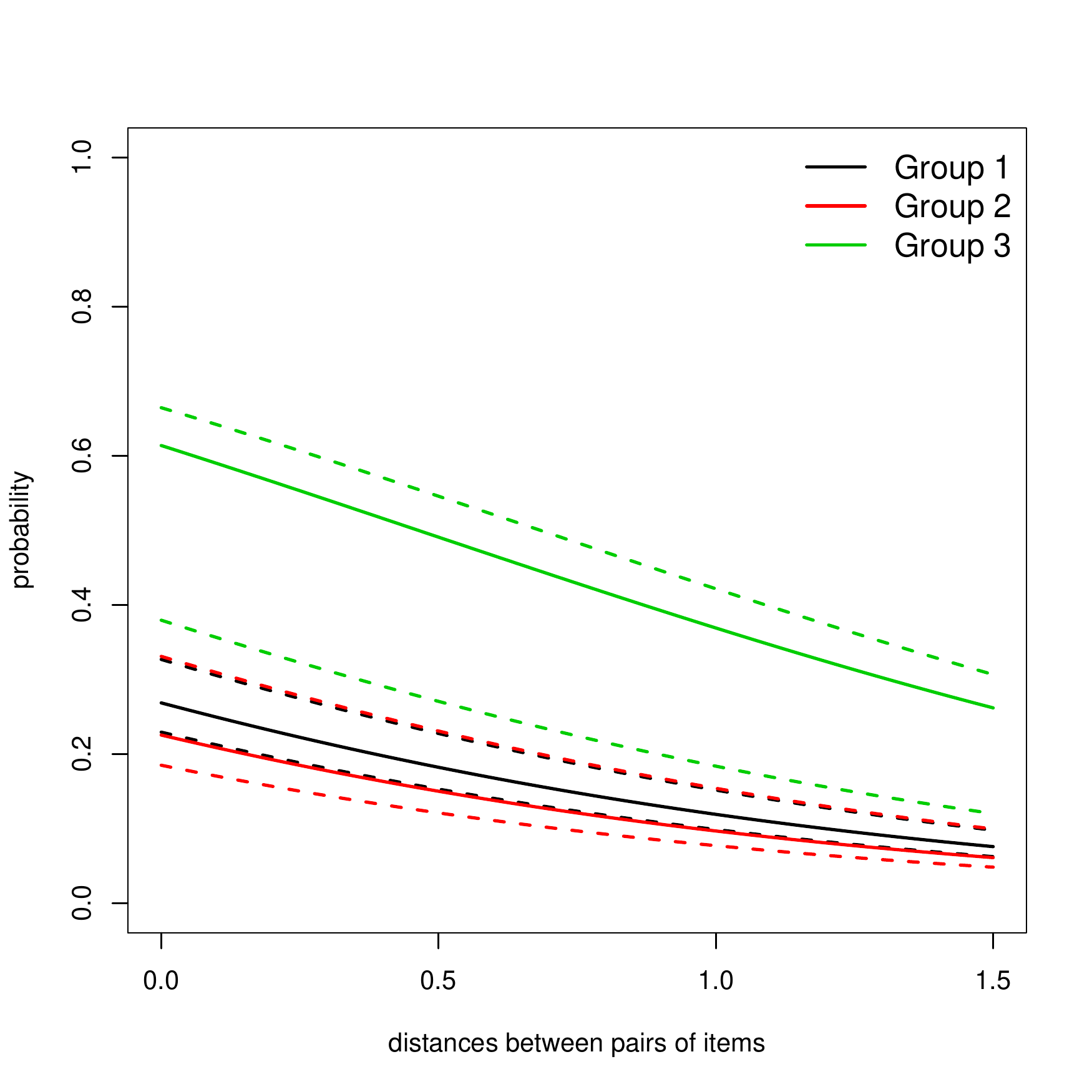} \\
\end{tabular}
\caption{
\label{fig:latent_drv1}
(a) A latent space for respondents with spectral clustering results. 
There are 3 respondent clusters - black circles represents Cluster 1; red circles represents Cluster 2; and green circles represents Cluster 3.
Note that item latent positions are added in this space with 4 item clusters (described in the legend);
and (b) Correct response probability functions for three respondent clusters
}
\end{figure}

\begin{center}
[Insert Figure \ref{fig:latent_drv1}]
\end{center}

Figure \ref{fig:latent_drv1}(a) displays the student clustering result in the two-dimensional Euclidean latent space.
Item latent positions and their clustering results are also added in this space,
so that we can more clearly understand the characteristics of 
the respondent clusters.
We observed that the students whose total score is close to the maximum score
are located close to the origin (prior mean of ${\bf Z}$),
whereas the students whose total score is close to the minimum (non-zero) score
are located relatively further away from the origin.
This affirms our earlier claim that different jumping rules should be applied to ${\bf Z}$ based on the respondents' total scores.

From the person clustering result in Figure \ref{fig:latent_drv1}(a), we found that
Cluster 3 students (green circles) tend to be located around the origin (prior mean) of the latent space,
meaning that these students provided correct answers more often than the students
who are clustered further way from the origin. 
This implies that students in Cluster 3 are most likely to 
correspond to the formal-operational stage.

We noted that although there is a clear position differentiation between Cluster 1 and Cluster 2 (i.e., Cluster 1 is located bottom right (black circles)
and Cluster 2 is located up left (red circles)),
based on the location difference only, it is difficult to determine which cluster corresponds to the concrete-operational or the transition stage.
To more clearly interpret these clusters, an in-depth examination would be needed on the response patterns of the students in Clusters 1 and 2.
This inspection revealed that students in Cluster 1 tend to correctly answer concrete items (MT/MP), while failing to give correct answers
to most of the logical fallacy items (NA/AC).
Cluster 2 students showed the opposite tendency; they tend to correctly answer fallacy items (NA/AC),
while sometimes giving incorrect answers to concrete items (MT/MP).
The item latent positions added on the respondent latent space (in Figure \ref{fig:latent_drv1}(a))
also helps us to more clearly understand what is going on.
The latent space shows that all concrete items (MT/MP) are located near person Cluster 1, whereas all fallacy items (NA/AC)  reside near Cluster 2.
Based on this observation, we concluded that
Cluster 1 students are likely to correspond  to the concrete-operational stage,
while Cluster 2 students are likely to correspond to the transition between the concrete- and formal-operational stages.

Figure  \ref{fig:latent_drv1}(b) displays
the probability of a pair of items being correctly answered in each person cluster (based on the spectral clustering results)
as a function of the latent space distance between the item pairs. A solid line represents the inverse logit of the mean $\theta$ value for each cluster.
Lower and upper dotted lines around each solid line indicate the inverse logit of the 25\% and 75\% quantile of the $\theta$ values in each group.
Black lines correspond to Cluster 1, red lines to Cluster 2, and green lines to Cluster 3.
From the figure, Cluster 3 students show clearly higher correct responses probabilities
than the students in other clusters. This confirms our conclusion that
Cluster 3 students correspond to the formal-operational group.
Clusters 1 and 2 students, however, showed little differences in their correct response probabilities.

In Figure  \ref{fig:latent_drv1}(b), we observe that as the latent space distances increase
(or the characteristics of the pair of persons become more different from each other),
the correct response probabilities decrease.
Note that the curves described in Figure  \ref{fig:latent_drv1}(b) are conceptually
similar to the person characteristic curves that are reported in regular IRT analysis;
the key difference is that we place the probability of a pair of items being correctly answered as a function of
the latent space distances between the pair of items.
The person characteristic curves describe a person's probability of correctly answering an item as
a function of the difficulty level of the item.

\subsubsection{Item Dependence Structure}

We first examined the item intercept parameter  ($\beta$) estimates.
All estimates and their 95\% HPD intervals are provided in Section A.1 of the supplementary material.
The result suggests that the item with MP, AB, and NE features and the item with MP, AB, and NN features
have the highest $\beta$ estimates,
meaning that these two item types are most likely to be correctly answered by the students (i.e., they are the easiest items) regardless of the students' ability levels.
On the other hand, the item with AC, AB, and NN features and the item with AC, CF, and NN features
show the smallest $\beta$ estimates among the 24 items,
meaning that these two items are the most difficult items to answer correctly.

Interestingly, when the AB feature was combined with the AC feature, the items became more difficult,
whereas when the AB feature was combined with the MP feature, the items became easier.
This result implies that there are likely to be
interaction effects between the type of inference and the content of conditional factors on the item difficulty.
Interactions between test design factors are likely to generate
local dependence among the items that share the same design factors.
In other words, a regular IRT model based on the local independence assumption
is likely to be inappropriate for DRV data analysis.



\begin{figure}[htbp]
\centering
\begin{tabular}{cc}
(a)  &
(b)  \\
\includegraphics[width=0.45\textwidth]{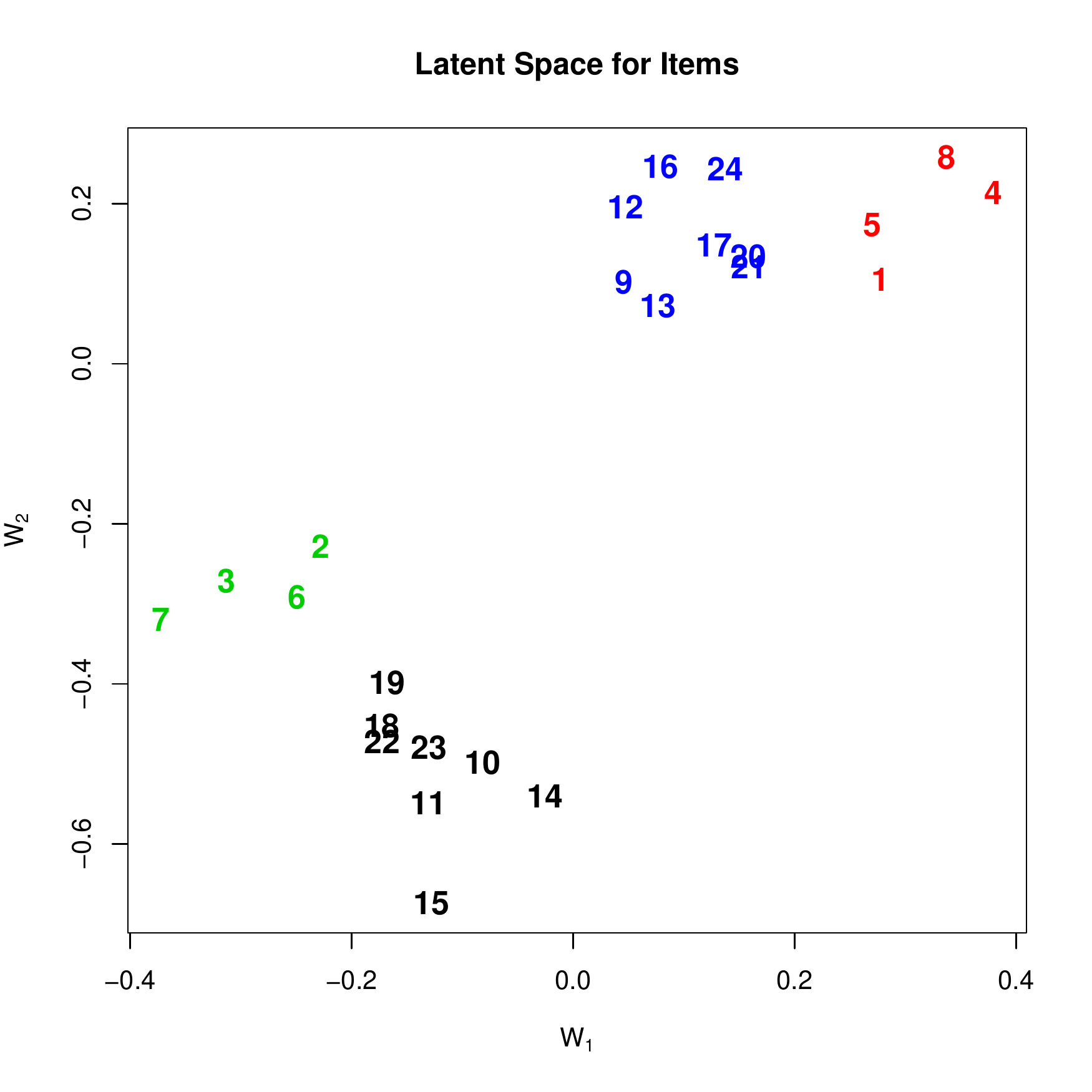} &
\includegraphics[width=0.45\textwidth]{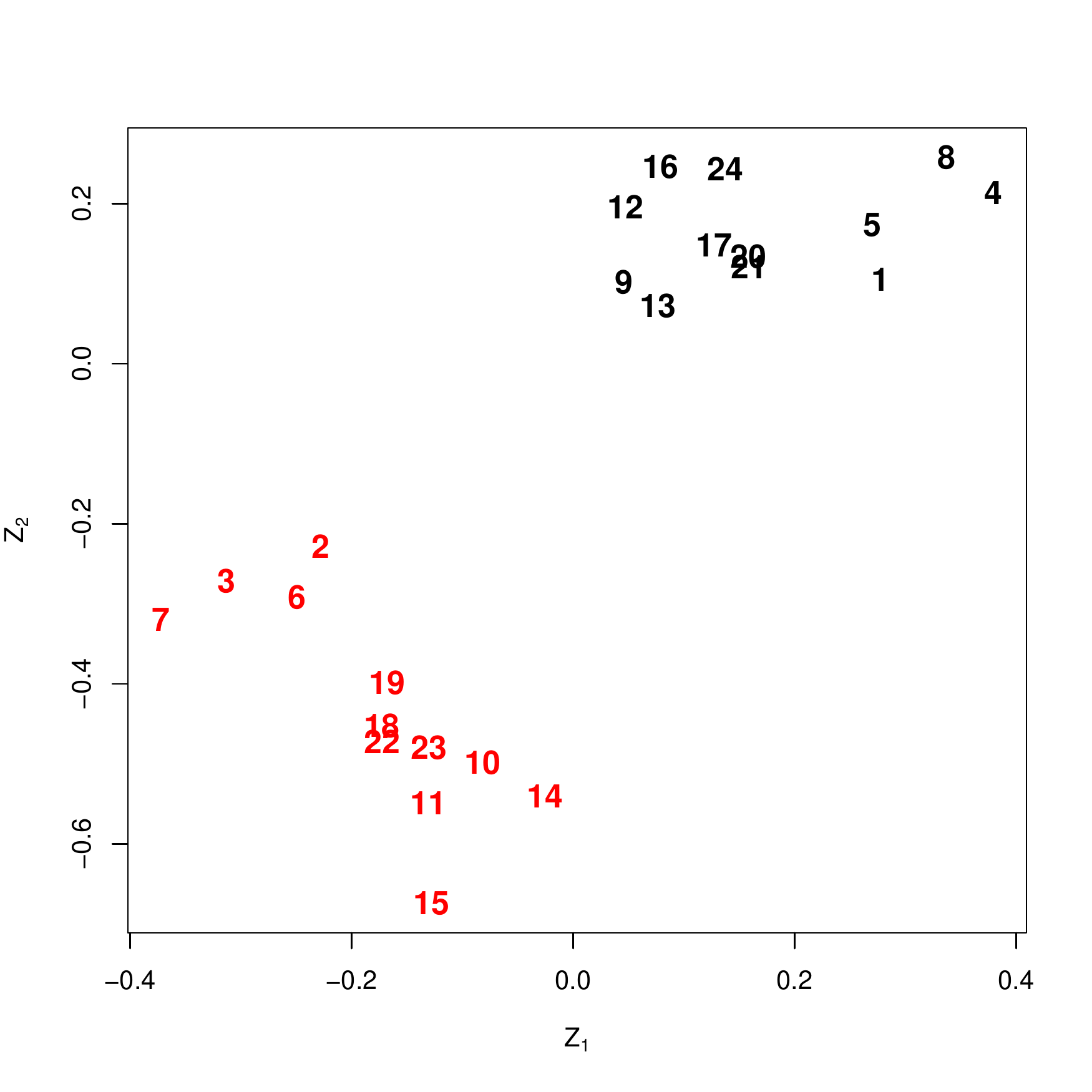} \\
(c)  &
(d)  \\
\includegraphics[width=0.45\textwidth]{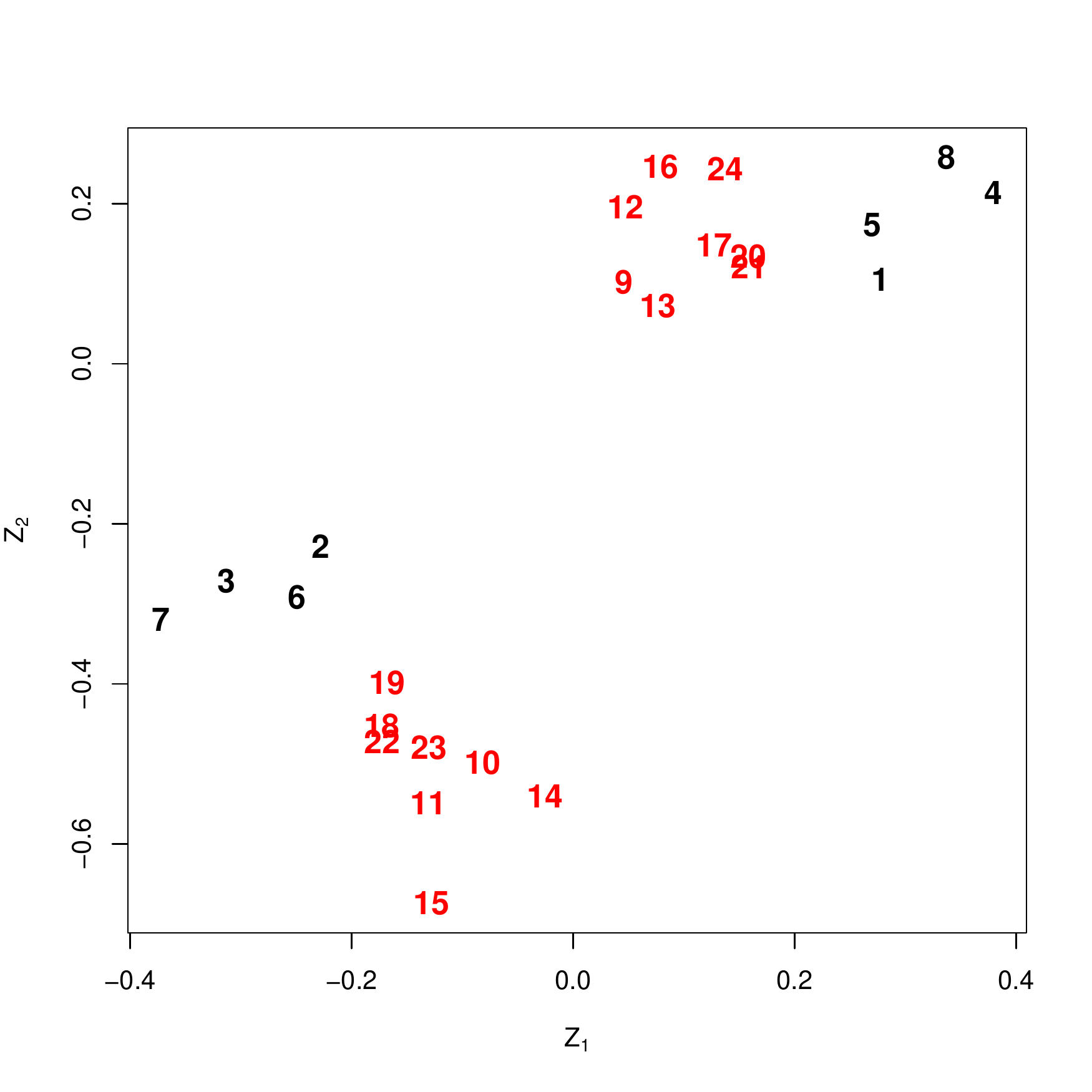} &
\includegraphics[width=0.45\textwidth]{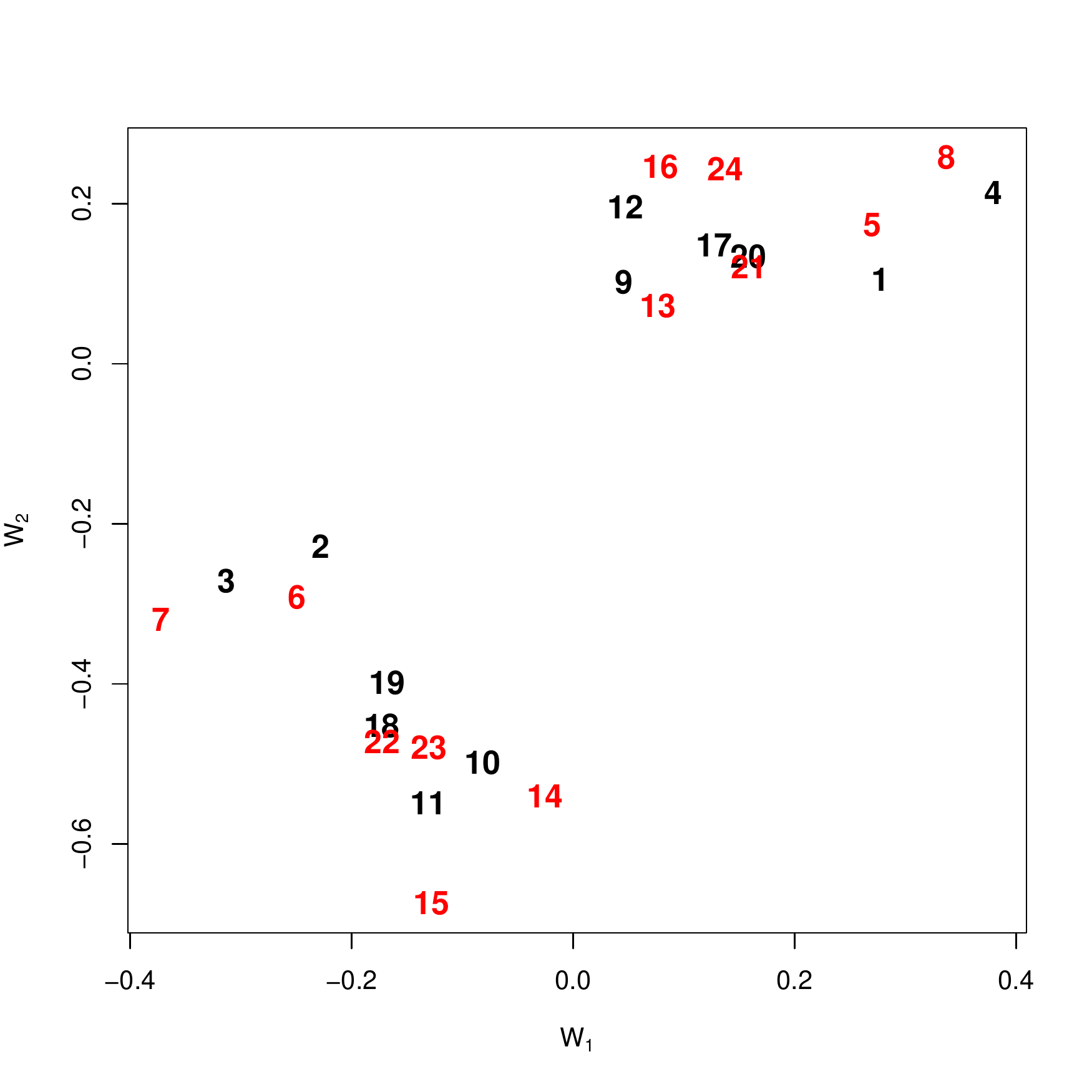} \\
\end{tabular}
\caption{\label{fig:latent_drv2}
Item latent spaces: (a) a latent space with spectral clustering results (with 4 clusters),
(b) a latent space for items color-coded by Type of inference,
(c) a latent space for items color-coded by Contents of the conditional, and
(d) a latent space for items color-coded by Presentation of antecedents.
}
\end{figure}

We then applied spectral clustering to the distance measures among the 24 items.
For item clustering, $k = 2$ was chosen which shows approximately 97.8\% of the variance being explained.
We then specified four clusters to describe the item dependence structure based on the visual clustering result shown in the latent space.
Figure \ref{fig:latent_drv2}(a) displays the spectral clustering results in the two-dimensional Euclidean latent space.

\begin{center}
[Insert Figure \ref{fig:latent_drv2}]
\end{center}

To aid in the visualization of an item dependent structure,
we assigned an item number to each item in Figure \ref{fig:latent_drv2}.
Specific descriptions on the item numbers used in Figure \ref{fig:latent_drv2}
and their corresponding design factors are provided in Section C of the supplementary material.

In Figure \ref{fig:latent_drv2}(a), we found that Cluster 1 consists of items that belong to a NA/AC group in Type of inference
and a AB/CF group in Content of the conditional.
Cluster 2 includes  items that belong to a MP/MT group in Type of inference
and a CO group in Content of the conditional.
Cluster 3 has  items that belong to a NA/AC group in Type of inference
and a CO group in Content of the conditional.
Cluster 4 contains  items that belong to a MP/MT group in Type of inference
and a AB/CF group in Content of the conditional. 
Note that the positions of Cluster 1 and Cluster 3 are roughly the reflections of Cluster 4 and Cluster 2 positions, 
respectively in the two-dimensional latent space, suggesting that items in Clusters 1 and 3 (NA/AC item group) 
have the opposite characteristics to the items in Clusters 4 and 2 (MP/MT item group),  respectively.  


Figures \ref{fig:latent_drv2}(b) to \ref{fig:latent_drv2}(d) color-code the items based on their design features in the two-dimensional item latent space.
Figure \ref{fig:latent_drv2}(b) is based on Type of inference;  the red items correspond to the NA/AC item group,
while the black items indicate the MP/MT item group.
Figure \ref{fig:latent_drv2}(c) is based on Content of the conditional; the red items correspond to the AB/CF item group,
while the black items indicate the CO item group.
Figure \ref{fig:latent_drv2}(d) is based on Presentation of antecedents; the red items correspond to the NE item group,
while the black items indicate the NN item group.
It is interesting to observe
in Figure \ref{fig:latent_drv2}(d) that the two Presentation of antecedents groups
are not clearly differentiated in the item latent space.
This means that the Presentation of antecedents design factor may not contribute to creating dependence among the test items.

\begin{figure}[htbp]
\centering
\begin{tabular}{cc}
(a) Cluster 1 &
(b) Cluster 2 \\
\includegraphics[width=0.45\textwidth]{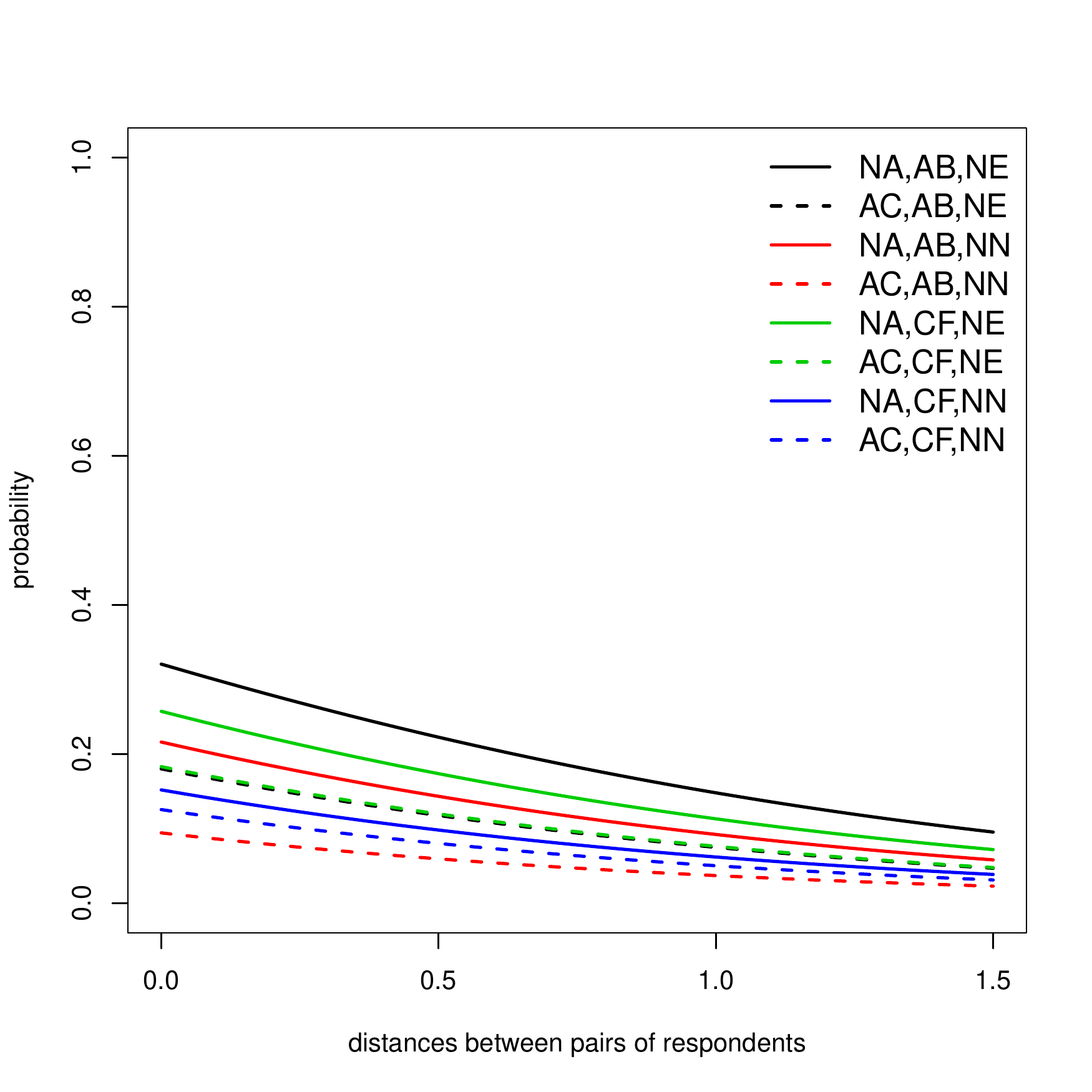} &
\includegraphics[width=0.45\textwidth]{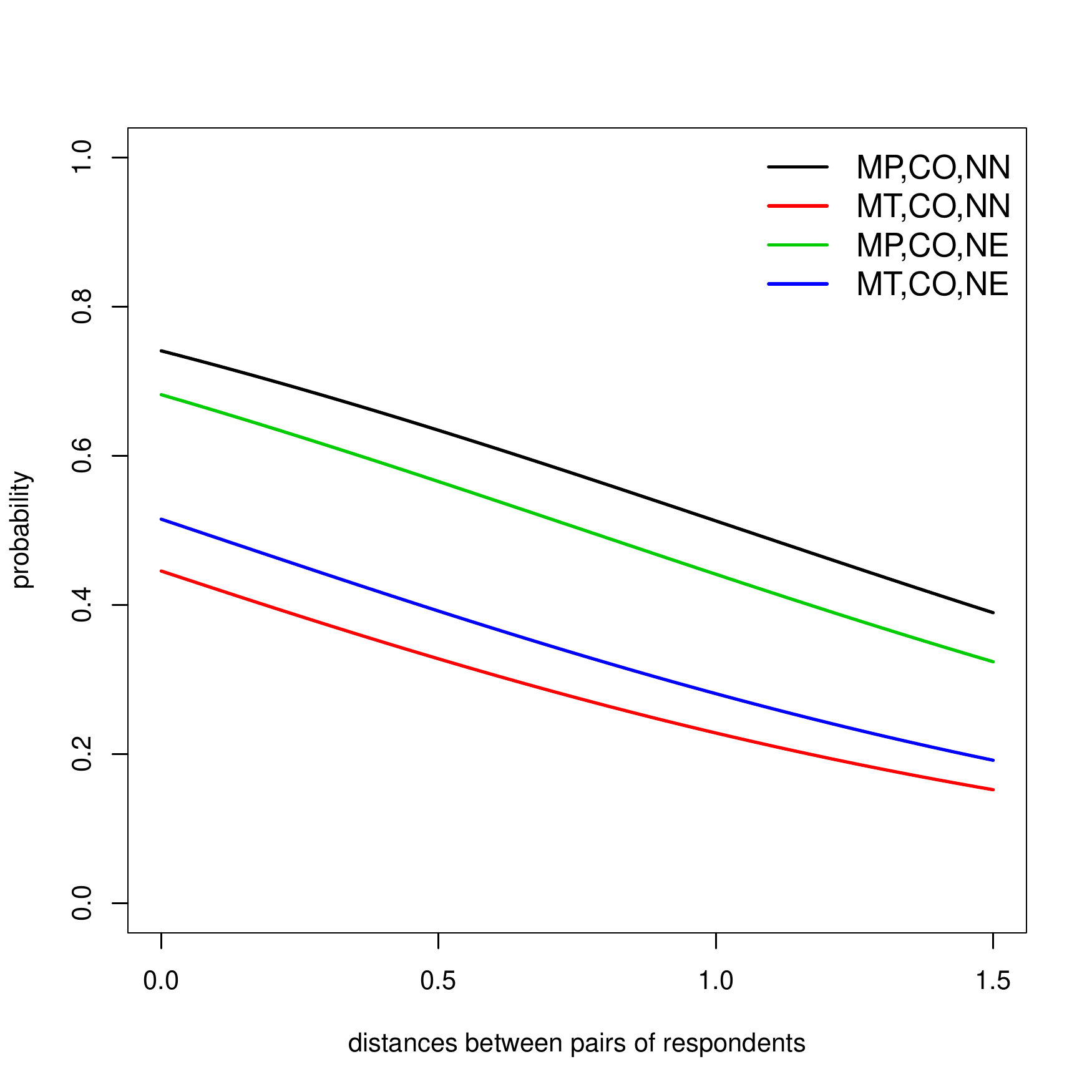} \\
(a) Cluster 3 &
(b) Cluster 4 \\
\includegraphics[width=0.45\textwidth]{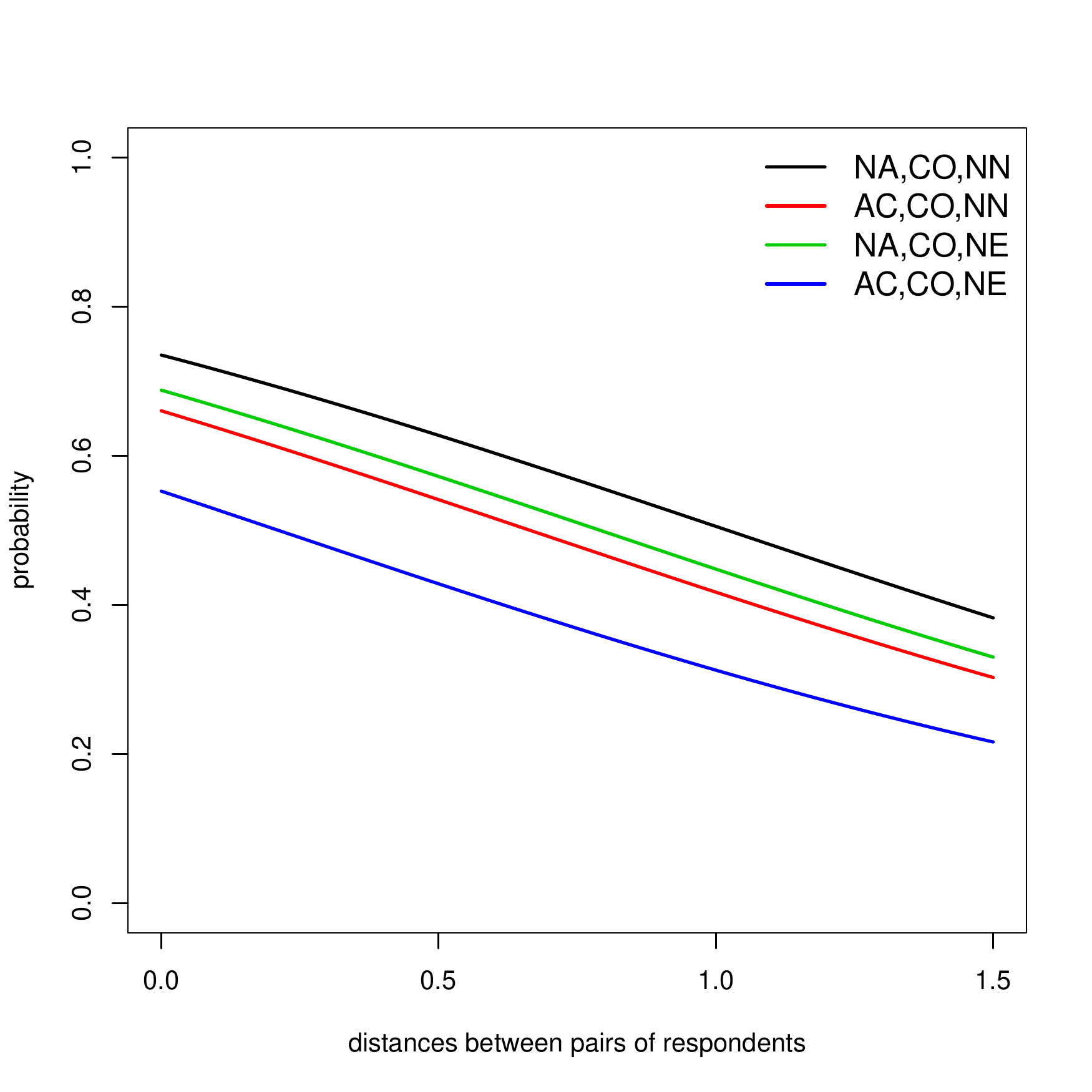} &
\includegraphics[width=0.45\textwidth]{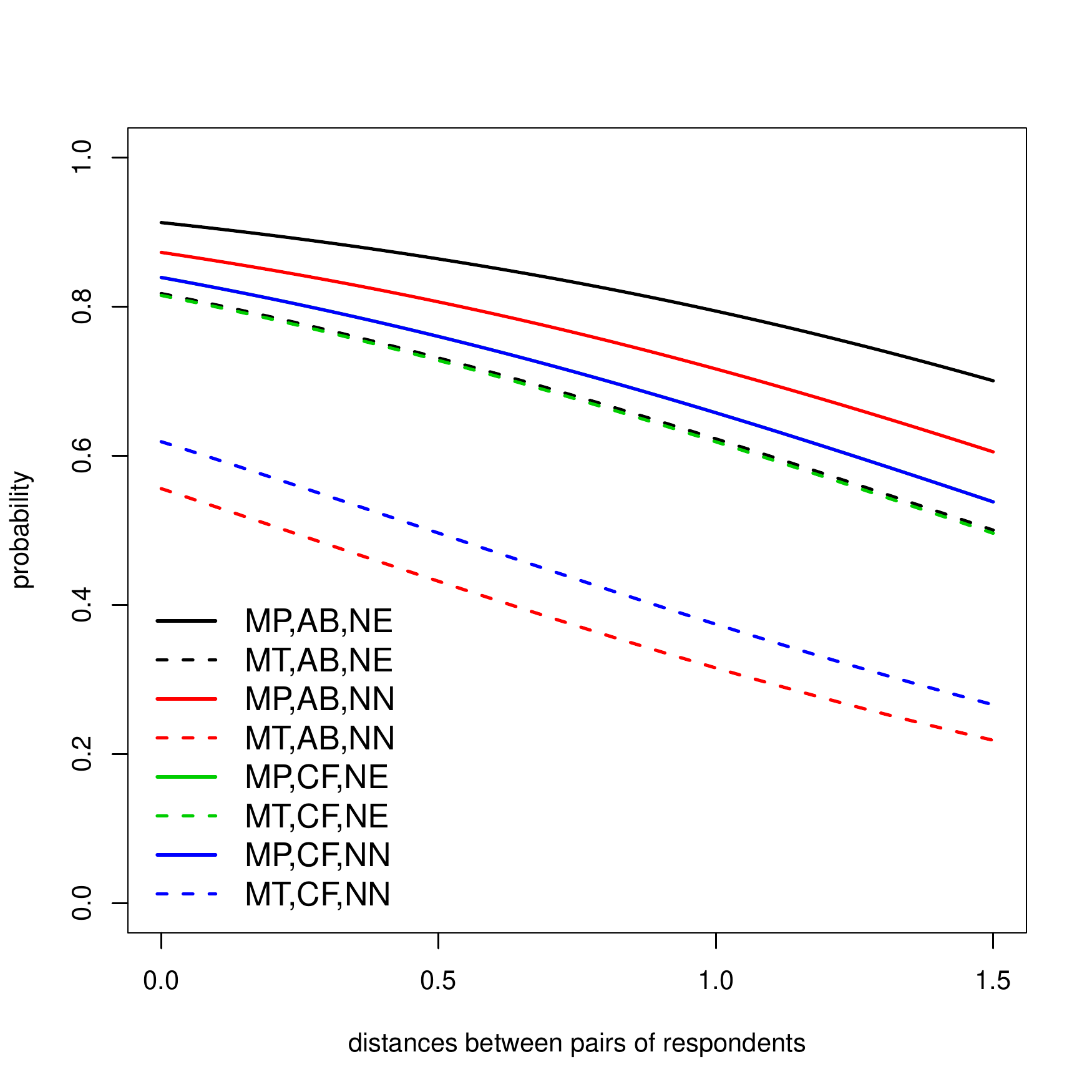} \\
\end{tabular}
\caption{\label{fig:latent_drv3}
Correct response probability functions for each of four item clusters
}
\end{figure}

\begin{center}
[Insert Figure \ref{fig:latent_drv3}]
\end{center}

Figure  \ref{fig:latent_drv3} displays
the probability of a pair of students correctly answering four item clusters (based on the spectral clustering results)
as a function of the latent space distance between the student pairs.
To summarize results, we found that
(1) NN item are generally easier than NE items,
(2) for CO items, the correct response probabilities are in the order of MP $\approx$ NA $>$ AC $>$ MT, and
(3) for AB/CF items, the correct response probabilities are in the order of MP $>$ MT $\gg$ NA $>$ AC.
That is, the difficulty order of the type of inference items is not consistent across the items' contents of the conditional features.

Note that the curves in Figure  \ref{fig:latent_drv3}
are similar to the item characteristic curves that are often reported in regular IRT analysis.
The difference is that we describe the probability of a pair of respondents correctly answering an item group
as a function of the latent space distances between the pair of respondents,
while a regular item characteristics curve describes the probability of a person's correctly answering an item
as a function of the person's ability level.

\subsection{Comparison with Mixture IRT Analysis}

A mixture Rasch model \citep{rost:90} is an extension of a regular Rasch model with an additional
categorical latent variable that is introduced to capture heterogeneity among respondents.
An important feature of a mixture Rasch model is that each latent class is allowed to
have a different measurement structure, i.e., a different set of item parameters.
A mixture Rasch model typically assumes that the respondents (or their latent traits)
follows a finite-mixture normal distribution, where the number of mixtures (or latent classes) is unknown a priori but empirically determined. 
And individual person's latent class membership is determined, e.g., using maximal a posteriori estimation.
Prior studies using the DRV data applied a finite-mixture Rasch model to analyze DRV data \citep{spiel:01}.



Note that mixture IRT analysis assumes independence of respondents within each cluster
as well as local independence of items given a person's ability \citep{Gollini:14}.
In the previous section, however, we showed that
there were non-negligible dependencies among items as well as among respondents.
This implies that the assumptions required by mixture Rasch analysis are likely to be violated with the DRV data.

The goal of this section is to compare DLSJM with a mixture Rasch approach
to examine whether ignoring or accounting for item and person dependencies in the data
would lead to different results and inferences.
The DLSJM and mixture Rasch models have different parameter structures;
thus, a direct comparison between the two approaches may be infeasible.
Hence, we focus on evaluating the performance of the two approaches
in terms of their person clustering solutions.

We applied a mixture Rasch model with three clusters to the DRV data using maximum likelihood estimation, similar to \citet{spiel:01}.
The most likely latent class for individual students was obtained using maximal a posteriori estimation.
We then compared the individual students' predicted latent class membership
with our spectral person clustering results.

\begin{center}
[Insert Table \ref{tab:mismatch}]
\end{center}

Overall, 57.2\% of the students were assigned to the same clusters based on the two approaches.
Table \ref{tab:mismatch} summarizes the proportions of classification mismatches between the two approaches.
Most mismatches are found on the diagonal area of the table, 
which relates to the
classification of the students into the formal or transition stages.
Specifically, a 13.9\% of the students were classified into the transition stage with DLSJM clustering; 
the students were classified to the formal-operational stage 
with mixture IRT (Case 1).
In addition, approximately 22.5\% of the students were classified into the formal-operational stage with DLSJM clustering; 
the students to the transition stage with mixture IRT (Case 2).

For a better understanding of the mismatch between the two classification solutions,
we examined the response patterns of the students in the the above two cases that displayed the most mismatches
(the response patterns of all cases in Table \ref{tab:mismatch}
are provided in Section D of the supplementary material).
In Case 1, we found that students tend to give correct answers to logical fallacy items (NA/AC)
but often give incorrect answers to a number of concrete items (MP/MT), regardless of the contents of the conditional.
Hence, it would be more reasonable to classify those students into the transition group rather than to the formal-operational group.
In Case 2, the students tend to correctly answer most of the concrete items with any conditional;
they give correct answers to the logical fallacy items that are combined with concrete or abstract conditional,
but fail to give correct answers to the local fallacy items that are combined with counterfactual conditional,
which are supposedly the most challenging test items.
Thus, it would be more sensible to assign these students to the formal-operational stage, rather than to the transition stage.

Based on this result, 
we concluded that DLSJM's person classification solution appears more reasonable than
the classification based on the mixture Rasch analysis. 
In particular, our classification was able to better 
distinguish the formal-operational versus the transition stage. 

\subsection{Comparison with Multidimensional IRT Analysis}

In this section, we compare the performance of DLSJM with an exploratory multidimensional IRT model 
in terms of item dependence structure detection. 
From the analysis performed in the previous section, DLSJM identified four item clusters from the DRV data - Cluster 1: NA/AC-AB/CF group,  
Cluster 2: MP/MT-CO group,  Cluster 3: NA/AC-CO group, and Cluster 4: MP/MT-AB/CF group. The clustering structure was visualized in Figure \ref{fig:latent_drv2}(a). 
Due to close proximity of the clusters in the latent space, 
it may be reasonable to collapse Clusters 1 \& 3 (NA/AC items) and Clusters 2 \& 4 (MP/MT items), which results in a simpler dependence structure with two item clusters. 



\begin{center}
[Insert Table \ref{tab:multidim}]
\end{center}

We conducted a traditional, exploratory multidimensional IRT analysis with the DRV data 
to see whether a similar item clustering structure would be identified with the traditional method. 
The {\tt mirt} package \citep{Chalmers:12} in R was employed for the exploratory IRT analysis 
with full information maximum likelihood estimation. 
We increased the number of dimensions from 2 to 6 (the six-factor model did not converge).  
The five-factor solution showed the best fit to the data based on 
model fit statistics (RMSEA=0.04, TLI=0.98, CFI=0.99), but its incremental model fit gain was minimal compared to the four-factor model (RMSEA=0.04, TLI=0.97, CFI=0.98) and the five-factor model's factor interpretation was challenging for the data.   
Hence, we selected the four-factor model to obtain the final item cluster solution (of the exploratory IRT analysis) and compared it with the four-cluster solution obtained from DLSJM. 
In addition, we also compared the two-factor solution from the exploratory IRT analysis with the simple structure (with two item clusters) identified with DLSJM.  
Table \ref{tab:multidim} presents the factor loading structures obtained from the two- and four-factor exploratory IRT analysis. 

The result shows that 
the item clustering result from the two-factor analysis is quite similar to the DLSJM's simple structure solution, 
with the exception of Item 16 (that corresponds to MT in Type of inference, CO in Content of the Conditional, and NE in Presentation of the antecedent).
Interestingly, however, the four-factor solution showed inconsistent results compared with DLSJM.  Specifically, it was found that 
Dimension 1 includes the items that belong to the MP/MT group in Type of inference and the CO/CF group in Content of the Conditional;
Dimension 2 contains the items that belong to the NA/AC group in Type of inference and the CF group in Content of the Conditional;
Dimension 3 contains the items that belong to the NA/AC group in Type of inference and the CO group in Content of the Conditional 
(with the exception of Item 11 that corresponds to AC in Type of inference, AB in Content of the Conditional, and NN in Presentation of the antecedent); and 
Dimension 4 includes the items that belong to the AB group in Content of the Conditional.

This solution from the exploratory IRT analysis appears unsatisfactory because (1) the CO and CF item groups, which are clearly different in terms of complexity levels, were not separated out when combined with MP/MT item features, and (2) the AB item group behaved independently of the CF item group, which is not in alignment with theoretical expectations. 
Based on these results, we concluded that 
the performance of the traditional, exploratory multidimensional IRT method, 
was less satisfactory in terms of detecting the complex item dependence structure of the DRV assessment data, when compared against the proposed DLSJM approach. 

\subsection{Simulation Study}


In this section, we provide a simulation study to further evaluate the DLSJM performance.
We chose a finite mixture IRT model as a comparison method (as in the empirical study section)
so that the DLSJM's clustering ability (person clustering) could be assessed relative to the performance of the existing method.\footnote{
None of the existing methods in psychometrics can identify both item and person clusters simultaneously; hence, a full scope comparison with an existing method may be infeasible.}

For simulations, we considered a situation that is analogous to the DRV data in terms of the data size (24 items and 300 students)
as well as in terms of the person and item dependence structures.
This way, we could not only evaluate the DLSJM's performance  but also validate the DLSJM's empirical analysis results
that we presented in the previous section.

We considered 6 item groups (with 4 items per group) and three person classes (with 100 respondents per class), while
setting the performance on item groups to be representative of one of the three person classes
(similar to concrete-operational, formal-operational, and transitional stages).
Specifically, we assumed that (1) respondents in Class 1 (the concrete-operational stage)
tend to give correct answers to item group 1 (item 1-4) and item group 2 (item 5-8);
(2) respondents in Class 2 (the transitional stage) tend to give accurate responses to item group 5 (item 17-20) and item group 6 (item 21-24); and
(3) respondents in Class 2 (the formal-operation stage) tend to give exact answers to item group 3 (item 9-12) and item group 4 (item 13-16).

Item responses are generated based on the following procedure:

\begin{enumerate}
\item[] [\textbf{Step 1: Extra person dependence generation}]
In the first step, we generate additional dependence among respondents within each of three classes.
Recall that each respondent class is expected to perform well (or to be representative of) with specific item groups
(e.g., respondents in Class 1 is expected to do well in item groups 1 and 2).
However, this kind of perfect situation may be unrealistic. For instance, some students in Class 1 may also perform well with item group 3,
while other students in Class 1 may perform poorly even with item group 1.
Note that such violations  create randomness (or extra dependencies) among some respondents within a class.
That is, our intention for including this step was to purposefully create randomness (or dependencies) in the data so that
the generated data could better reflect more realistic 
assessment scenarios. 

To this purpose, we consider actual ``inside-class'' and ``outside-class'' probabilities for item groups,
so that for respondent $k$ ($=1,\cdots,100$) in Class $m$ ($=1,\cdots,3$), item group $g_i$ ($=1,\cdots,6$)
can actually be assigned to either ``inside'' or ``outside'' of the intended class with some probabilities.
For instance, for respondents 1-100 whose intended class is Class 1 (Concrete-operational stage),
item groups 1 and 2 are ``intended inside-class'' item groups, while  item groups 3 to 6 are ``intended outside-class'' item groups. We then
 \begin{itemize}
 \item Assign $g_i$ to ``actual inside-class'' with probability $p_{11}$ if $g_i$ are ``intended inside-class'' item groups,
 \item Assign $g_i$ to ``actual outside-class'' with probability $p_{21}$ if $g_i$ are ``intended outside-class'' item groups.
 \end{itemize}

That is, we allow for item groups in ``intended inside-class'' to be ``actual inside-class'' or be ``actual outside-class'' item groups
(similarly, item groups in ``intended outside-class'' to be ``actual outside-class'' or be ``actual inside-class'' item groups).
This way, when item groups 1 and 2 are ``intended inside-class'' item groups (for Class 1 respondents),
for instance, some respondents in this class can perform well with item groups other than item groups 1 and 2 and/or perform poorly with item groups 1 or 2.
To give a more concrete example, let us return to the DRV assessment, where  
children in the concrete-operational stage are expected to correctly answer NA/AC items but incorrectly answer MT/MP items. This theoretical expectation, however, may not hold in reality. 
To infuse some dose of reality (or noise) into the data generation process, we can utilize the ``inside-class''  to make children in the concrete-operational stage to periodically answer MT/MP items  incorrectly, while utilizing the ``outside-class'' probabilities to account for that children in the concrete-operational stage to periodically answer NA/AC items correctly.

We set the ``inside-class'' probabilities to always be higher than the ``outside-class'' probabilities (while both probabilities are less than 1)
so that we make sure only small degree of randomness (or extra dependencies) is created among respondents within a class.


\vspace{.2 in}
\item[] [\textbf{Step 2: Item response generation}]
In the second step, item responses are generated within item group $g_i$ as follows:
 \begin{itemize}
 \item For item group $g_i$ that belongs to the ``actual inside-class'', generate binary responses with probability $p_{12}$.
 \item For item  group $g_i$  that belongs to the ``actual outside-class'', generate binary responses with probability $p_{22}$.
 \end{itemize}

\item[] [\textbf{Step 3: Item dependence generation}]
In the third step, local item dependence is generated within item group $g_i$.
To this purpose, we followed the procedure described by \citet{chen:97}:
\[ \begin{split}
 \mbox{With probability $1 - \rho$, } X_{.j} &=
 \left\{\begin{array}{cc} 1, & \mbox{with }P(X_{.j}=1 \mid \theta)\\ 0, & \mbox{with }P(X_{.j}=0 \mid \theta). \end{array}\right.\\
 \mbox{With probability $\rho$, } X_{.j} &=
 \left\{\begin{array}{cc} 1, & \mbox{with }X_{.i}=1\\ 0, & \mbox{with }X_{.i}=0. \end{array}\right.
\end{split} \]

That is, we set up a situation where a test taker's response to item $j$ is affected by his/her response to item $i$ (when $i \neq j$) with probability $\rho$.
Suppose $\rho = 0.7$. In this case, a person 
has a 70\% chance of providing a correct response to item $j$ 
 if he/she gave a correct response to item $i$.
On the other hand, he/she has a 30\% chance of giving a correct or incorrect response to item $i$ based on his/her ability level and the difficulty of the item.
Hence, as $\rho$ increases, local dependence between the two pairs of items increases.
\citet{chen:97} labeled this type of local dependence as \emph{surface local dependence} and explained that surface local dependence arises,
for instance, in a test with similar items (in content or location in the test).
A pair of items is very similar; thus, the test taker responds identically to those items.


\end{enumerate}

We considered six sets of the actual ``inside-class'' probabilities  with $p_{11} = (0.7, 0.8, 0.9)$ and $p_{12} = (0.7, 0.8)$,
while fixing the actual ``outside-class'' probabilities to $(p_{21}, p_{22}) = (0.5, 0.5)$ and the item dependence probability to $\rho=0.8$.
Thereby, we have a total of six data generation conditions.
Under each condition, we generated 200 datasets
and applied the DLSJM and the mixture Rasch model (with three latent classes).
For both methods, the same estimation setting was applied as in the DRV empirical data analysis.
We then compared the person classification results from the DLSJM's spectral clustering (DLSJM Clustering) with the mixture Rasch analysis (Mixture-Rasch).

Table \ref{tab:sim}
lists DLSJM's and mixture Rasch model' average probabilities of classifying the students into each cluster per simulation condition.
Classification accuracy is summarized in a 3 $\times$ 3 matrix per condition;
the row represents true clusters, while the column indicates the predicted clusters.
The diagonal elements in bold represent correct classification probabilities.

\begin{center}
[Insert Table \ref{tab:sim}]
\end{center}

In summary, the DLSJM analysis with spectral clustering shows approximately 10-20\% higher correct classification  probabilities
than the mixture Rasch analysis in all simulation settings.
Specifically, DLSJM spectral clustering shows .70 to .79 correct classification probabilities for Cluster 1,
.71 to .78 for Cluster 2, and .76 to .79 for Cluster 3, whereas mixture Rasch analysis shows .50 to .57 for Cluster 1,
.50 to .57 for Cluster 2, and .57 to .76 for Cluster 3.
This result suggests that when there is a complex dependence structure among respondents and among items (similar to the DRV data),
DLSJM outperforms the mixture Rasch approach in terms of person clustering.

\section{Discussion}

In this paper, we proposed a latent space joint model to analyze item response data. Unlike regular IRT models that are often utilized for item response data analysis, 
our proposed model does not require  local item and person independence assumptions. Furthermore, our approach provides relative distances between
pairs of items, which can be used to identify item dependence or item cluster structures in two-dimensional Euclidian latent spaces.

Our approach began with constructing two 
latent space joint models 
for multiple layers of person networks and item networks.
We then combined the two latent space joint models
by viewing the latent space for items as a function of latent spaces for people.
We labeled the resulting, integrated model as a doubly latent space joint model (DLSJM)
and provided a fully Bayesian approach for estimating the model.

DLSJM is a unique analytic tool for psychometricians.  Our DLSJM method can be utilized to simultaneously identify unknown 
potentially convoluted item and person dependence (or cluster) structures in item response data. 
As discussed earlier,  DLSJM is more advantageous than traditional methods, such as finite mixture IRT analysis and exploratory IRT analysis, which can identify only person or item clusters.  
Another merit of DLSJM, when compared with non-model-based methods such as MDS and DETECT, is that DLSJM includes item and person parameters which may be conceptually similar to the parameters of standard one-parameter IRT models. 
Specifically, DLSJM's item parameters indicate the probability of correctly answering the item when a pair of respondents has the same ability levels. 
The person parameters indicate the respondent's probability of giving correct responses to a pair of items that has the same level of difficulty. 
Thus, the DLSJM's item and person parameters can be used to discuss and compare the overall easiness of individual items as well as the overall ability levels of individual respondents as in standard IRT analysis. 


We applied DLSJM to analyze the item response data from the DRV assessment that is developed based on a complex item design structure.  
We found that the test items that have AC and AB design features are 
the most difficult items, while the items that are associated with MP and AB design features
are the easiest items.
This result suggests that there seems to be interactions between the two test design factors, type of inference (AC, MP) and
the content of conditionals (AB).
This type of interactions between test design factors
is likely to generate dependencies among items.
Inspecting the item dependence structure in the latent space
confirmed that there was indeed a strong dependence structure among the items that show the two design features. 
We also applied an exploratory multidimensional IRT analysis to the data in order to examine whether this traditional method would be able to duplicate DLSJM's ability  to identify the item dependence structure. 
The results showed that the exploratory IRT method performed less satisfactorily than DLSJM 
in identifying the data's convoluted item clustering structure. 

Further, we applied a spectral clustering technique to visualize the person dependence structure in a latent space.
Our analysis provided  information on which of the three clusters
(that correspond to the concrete-operational stage, the formal-operational stage, and the transition between the stages)
each student of the DRV data belongs to.
For evaluating person clustering accuracy, we compared our results
with the results from a mixture Rasch analysis.
This comparison showed that our approach was more effective 
than mixture Rasch analysis
in distinguishing the transition stage from the formal-operational stage.
In a simulation study, we further confirmed that
our approach outperforms the mixture Rasch approach in terms of person classification accuracy.

Instead of using a spectral clustering technique,
a model-based clustering \citep{Handcock:2007} may be applied by assigning a Gaussian mixture model for ${\bf z}_k$.
In our analysis, we chose spectral clustering because of its computational efficiency.
It can be shown that both spectral and model-based clustering methods are mainly
based on pairwise distance measures.
Spectral clustering, however, is much simpler and faster than
model-based clustering that may require an extremely large number of MCMC iterations to improve mixing \citep{Handcock:2007}.

Since all networks that we constructed for our analysis contain only cliques (subnetworks where every edge is present),
we have a rather simplistic network structure with no heterogeneity.
Hence, one may argue that a simpler network approach may be as effective as a latent space modeling approach for item response analysis.
However, we adopted a latent space modeling framework because it allows for
easier interpretations of $\beta$ and $\theta$ (similar to IRT model parameters)
in that the latent space model is formulated within a logistic modeling framework,
similar to standard IRT models.
In addition, with item response data, it is necessary to manage multi-layer networks (for persons and items),
which can be easily handled with a latent space modeling approach.

Recently, network modeling approaches based on an Ising model for item response data
has been proposed in psychometrics \citep{vanBorkulo:2014, Kuris:2016, Epskamp:2017}.
One may think that their methods may be similar to our approach.
However, there are important differences: these methods 
 only  identify an item network (dependence) structure, but not a person network structure.
In addition, their approaches still require the local item independence assumption to avoid the computational difficulty that arises
from doubly-intractable normalizing constants of the Ising model.

Finally, we would like to add that
our approach could be generally applied to analyze other binary item response data.
More importantly, our method can also be applied to develop statistical indices to evaluate local dependencies among items and among people,
which can be useful for test construction and test validation.

\section*{Acknowledgement}

We appreciate the Editor, the Associate Editor, and three anonymous reviewers for their careful reading and detailed comments on the previous versions of our manuscript. 

\bibliography{reference}
\vspace{\fill}
\newpage

\centerline{Figures}

\medskip

\vspace{\fill}

\newpage

\vspace{\fill}
\newpage

\vspace{\fill}
\newpage

\newpage

\newpage

\newpage

\centerline{Tables}

\medskip
%
%

\begin{table}[htbp]
\centering
\begin{tabular}{cccc}
\hline
                   & concrete (Mixture) & transition (Mixture) & formal (Mixture) \\
\hline
concrete (DLSJM)   &  -  & 3.3 & 0.2  \\
transition (DLSJM) &  0   &   - & 13.9  \\
formal (DLSJM)     &  2.9   &  22.5    & - \\
\hline
\end{tabular}
\caption{\label{tab:mismatch}
Proportions of individuals who were assigned to different clusters with the mixture IRT approach (Mixture)
compared to spectral clustering of students' latent spaces, which were estimated from DLSJM (DLSJM).
Overall, approximately 43.5\% of students were assigned to different classes with the mixture IRT method.
}
\end{table}
\vspace{\fill}
\newpage

\begin{table}[htbp]
\centering
\begin{tabular}{c|cc|cccc}
\hline
		 & \multicolumn{2}{c|}{Dimension = 2} & \multicolumn{4}{c}{Dimension = 4} \\
         & Factor 1 & Factor 2 & Factor 1 & Factor 2 & Factor 3 & Factor 4 \\
\hline
 1 & {\bf  0.937} &  0.243 & {\bf  0.640} &  0.424 & -0.352 & -0.127\\
 2 & -0.386 & {\bf  0.553} & -0.009 &  0.062 & {\bf  0.711} &  0.145\\
 3 & -0.434 & {\bf  0.589} & -0.081 &  0.086 & {\bf  0.894} & -0.073\\
 4 & {\bf  0.780} &  0.001 & {\bf  0.556} &  0.163 & -0.346 & -0.112\\
 5 & {\bf  0.811} &  0.092 & {\bf  0.551} &  0.307 & -0.384 & -0.115\\
 6 & -0.283 & {\bf  0.682} &  0.151 &  0.126 & {\bf  0.831} &  0.094\\
 7 & -0.465 & {\bf  0.570} & -0.102 &  0.053 & {\bf  0.925} & -0.070\\
 8 & {\bf  0.768} & -0.107 & {\bf  0.558} &  0.054 & -0.365 & -0.136\\
 9 & {\bf  0.420} & -0.202 &  0.171 &  0.109 &  0.133 & {\bf -0.808}\\
10 &  0.089 & {\bf  0.883} &  0.272 &  0.388 &  0.347 & {\bf  0.478}\\
11 &  0.036 & {\bf  0.833} &  0.220 &  0.331 & {\bf  0.416} &  0.390\\
12 & {\bf  0.393} & -0.335 &  0.171 & -0.012 &  0.094 & {\bf -0.829}\\
13 & {\bf  0.373} & -0.118 &  0.077 &  0.228 & -0.063 & {\bf -0.611}\\
14 &  0.068 & {\bf  0.775} &  0.217 &  0.327 &  0.185 & {\bf  0.535}\\
15 & -0.112 & {\bf  0.712} &  0.066 &  0.281 &  0.278 & {\bf  0.483}\\
16 &  0.149 & {\bf -0.416} & -0.016 & -0.088 & -0.040 & {\bf -0.607}\\
17 & {\bf  0.628} & -0.118 & {\bf  0.761} & -0.184 &  0.129 & -0.157\\
18 &  0.039 & {\bf  0.841} & -0.150 & {\bf  0.786} &  0.141 &  0.029\\
19 &  0.107 & {\bf  0.853} & -0.077 & {\bf  0.768} &  0.159 &  0.004\\
20 & {\bf  0.528} & -0.290 & {\bf  0.694} & -0.451 &  0.120 & -0.096\\
21 & {\bf  0.613} & -0.092 & {\bf  0.701} & -0.157 &  0.093 & -0.144\\
22 &  0.108 & {\bf  0.861} & -0.005 & {\bf  0.728} &  0.175 &  0.076\\
23 &  0.076 & {\bf  0.783} & -0.118 & {\bf  0.705} &  0.077 &  0.107\\
24 & {\bf  0.421} & -0.356 & {\bf  0.590} & -0.382 & -0.187 &  0.097\\
\hline
\end{tabular}
\caption{\label{tab:multidim}
Exploratory multidimensional IRT analysis results with 2 and 4 factor solutions for the DRV data. 
In each solution, the largest factor loading per item is marked in bold.  
}
\end{table}
\vspace{\fill}
\newpage

\begin{table}[ht]
\centering
\begin{tabular}{c|rrrr|rrrr}
Setting &  \multicolumn{4}{c|}{DLSJM Clustering} &  \multicolumn{4}{c}{Mixture-Rasch} \\ \hline
\multirow{ 4}{*}{$p_{11}$ = 0.7; $p_{12}$ = 0.7}
&   &    1 &    2 &    3 &   &   1 &    2 &    3   \\
& 1 & {\bf 0.70} & 0.21 & 0.09 & 1 & {\bf 0.50} & 0.33 & 0.17\\
& 2 & 0.20 & {\bf 0.71} & 0.09 & 2 & 0.33 & {\bf 0.50} & 0.17\\
& 3 & 0.12 & 0.12 & {\bf 0.76} & 3 & 0.20 & 0.14 & {\bf 0.66}\\ \hline
\multirow{ 4}{*}{$p_{11}$ = 0.7; $p_{12}$ = 0.8}
&   &    1 &    2 &    3 &   &   1 &    2 &    3   \\
& 1 & {\bf 0.75} & 0.18 & 0.07 & 1 & {\bf 0.51} & 0.35 & 0.14\\
& 2 & 0.19 & {\bf 0.74} & 0.07 & 2 & 0.34 & {\bf 0.51} & 0.15\\
& 3 & 0.11 & 0.10 & {\bf 0.79} & 3 & 0.18 & 0.25 & {\bf 0.57}\\ \hline
\multirow{ 4}{*}{$p_{11}$ = 0.8; $p_{12}$ = 0.7}
&   &    1 &    2 &    3 &   &   1 &    2 &    3   \\
& 1 & {\bf 0.74} & 0.17 & 0.09 & 1 & {\bf 0.52} & 0.33 & 0.14\\
& 2 & 0.18 & {\bf 0.73} & 0.09 & 2 & 0.33 & {\bf 0.52} & 0.14\\
& 3 & 0.12 & 0.12 & {\bf 0.77} & 3 & 0.17 & 0.14 & {\bf 0.68}\\ \hline
\multirow{ 4}{*}{$p_{11}$ = 0.8; $p_{12}$ = 0.7}
&   &    1 &    2 &    3 &   &   1 &    2 &    3   \\
& 1 & {\bf 0.75} & 0.18 & 0.07 & 1 & {\bf 0.54} & 0.32 & 0.14\\
& 2 & 0.17 & {\bf 0.75} & 0.07 & 2 & 0.34 & {\bf 0.54} & 0.12\\
& 3 & 0.12 & 0.11 & {\bf 0.78} & 3 & 0.15 & 0.21 & {\bf 0.64}\\ \hline
\multirow{ 4}{*}{$p_{11}$ = 0.9; $p_{12}$ = 0.7}
&   &    1 &    2 &    3 &   &   1 &    2 &    3   \\
& 1 & {\bf 0.76} & 0.16 & 0.08 & 1 & {\bf 0.55} & 0.33 & 0.12\\
& 2 & 0.16 & {\bf 0.75} & 0.09 & 2 & 0.32 & {\bf 0.55} & 0.13\\
& 3 & 0.12 & 0.12 & {\bf 0.76} & 3 & 0.14 & 0.10 & {\bf 0.76}\\ \hline
\multirow{ 4}{*}{$p_{11}$ = 0.9; $p_{12}$ = 0.7}
&   &    1 &    2 &    3 &   &   1 &    2 &    3   \\
& 1 & {\bf 0.78} & 0.15 & 0.07 & 1 & {\bf 0.57} & 0.32 & 0.11\\
& 2 & 0.15 & {\bf 0.78} & 0.07 & 2 & 0.33 & {\bf 0.57} & 0.10\\
& 3 & 0.11 & 0.11 & {\bf 0.79} & 3 & 0.13 & 0.22 & {\bf 0.65} \\
\hline
\end{tabular}
\caption{\label{tab:sim}
Summary of simulation study results.
}
\end{table}

\vspace{\fill}

\end{document}